\documentclass[twocolappendix,apj]{emulateapj}

\usepackage{perpage}
\usepackage{hyperref}
\usepackage{subdepth}
\usepackage{pdftexcmds}

\newcommand{\hms}[3]{#1^{\text{h}}#2^{\text{m}}#3^{\text{s}}}
\newcommand{\dms}[3]{#1^{\circ}#2'#3''}

\def\I{\,\textsc{i}}
\def\II{\,\textsc{ii}}

\def\HII{H\,\textsc{ii}}
\def\HI{H\,\textsc{i}}
\def\kms{\text{km s}^{-1}}

\shorttitle{CO SNR Survey}
\shortauthors{Kilpatrick, Bieging, \& Rieke}

\begin{document}


\title{A Systematic Survey for Broadened CO Emission Toward Galactic Supernova Remnants}

\author{Charles D. Kilpatrick, John H. Bieging, \& George H. Rieke}
\affil{Steward Observatory, University of Arizona, Tucson, AZ 85721}


\begin{abstract}

We present molecular spectroscopy toward 50 Galactic supernova remnants (SNRs) taken at millimeter wavelengths in $^{12}$CO and $^{13}$CO $J=2-1$ with the Heinrich Hertz Submillimeter Telescope as part of a systematic survey for broad molecular line (BML) regions indicative of interactions with molecular clouds (MCs).  These observations reveal BML regions toward nineteen SNRs, including nine newly identified BML regions associated with SNRs (G08.3$-$0.0, G09.9$-$0.8, G11.2$-$0.3, G12.2$+$0.3, G18.6$-$0.2, G23.6$+$0.3, 4C$-$04.71, G29.6$+$0.1, G32.4$+$0.1). The remaining ten SNRs with BML regions confirm previous evidence for MC interaction in most cases (G16.7$+$0.1, Kes 75, 3C 391, Kes 79, 3C 396, 3C 397, W49B, Cas A, IC 443), although we confirm that the BML region toward HB 3 is associated with the W3(OH) \HII\ region, not the SNR.  Based on the systemic velocity of each MC, molecular line diagnostics, and cloud morphology, we test whether these detections represent SNR-MC interactions.  One of the targets (G54.1$+$0.3) had previous indications of a BML region, but we did not detect broadened emission toward it.  Although broadened $^{12}$CO $J=2-1$ line emission should be detectable toward virtually all SNR-MC interactions we find relatively few examples; therefore, the number of interactions is low.  This result favors mechanisms other than SN feedback as the basic trigger for star formation.  In addition, we find no significant association between TeV gamma-ray sources and MC interactions, contrary to predictions that SNR-MC interfaces are the primary venues for cosmic ray acceleration.

\end{abstract}




\section{Introduction}\label{sec:intro}

Supernovae (SNe) inject kinetic energy, momentum, and enriched ejecta into the interstellar medium (ISM) and are thought to be one of the main sources of interstellar turbulence and galactic outflows.  An important factor in understanding these processes is the evolution of supernova remnants (SNRs) as their ejecta encounter various phases of the ISM.  For example, a large fraction of core-collapse SNe from high-mass stars may explode promptly and interact with the molecular cloud (MC) from which they formed.  This hypothesis suggests that the direct interaction between supernova (SN) ejecta and MCs, which we refer to as \textit{SNR-MC interactions}, may occur at a rate significantly higher than expected from the filling fraction of molecular gas in the ISM \citep{el77}.  While it has been estimated that as many as half of all SNRs have ejecta in physical contact with MCs \citep{rm01}, this estimate is highly uncertain and relies on a number of assumptions about massive stars, SNe, and the relation between SNe and MCs.  

Detection of SNR-MC interactions has traditionally relied on a number of methods for observing shocked molecular gas directly.  Strong, narrow emission from the OH $1720~\text{MHz}$ maser ($^{2} \Pi_{3/2}, J = 3/2, F = 2 \rightarrow 1$) is often detected in the vicinity of SNRs, especially those with other signatures of SNR-MC interactions \citep{gr68,hc77,den79a}.  More recent surveys have established that targeted searches for OH maser emission may be a reliable method of searching for SNR-MC interactions \citep[e.g.,][]{frail96,green97}.  To date, however, OH masers have only been detected toward $\sim 10\%$ of the 294 known SNRs in our galaxy \citep{green14}, although a lack  of detailed spectroscopic information and localization makes it impossible to associate all of these detections with SNRs.  This low incidence relative to the expected rate of SNR-MC interactions may be due to the unique physical conditions required for maser emission.  OH $1720~\text{MHz}$ masers are collisionally excited behind slow ($< 45~\kms$) C-type shocks in regions with moderate temperatures and densities ($T \sim 50 - 125~\text{K}$, $n_{\text{H}_{2}} \sim 10^{5}~\text{cm}^{-3}$) \citep{lockett99}.  Studies of the warm, diffuse H$_{2}$ regions around interacting SNRs indicate that physical conditions are often outside this range \citep{hrar09}.  Regions with higher shock velocities could also depopulate the hydroxyl $^{2} \Pi_{3/2}$ state or dissociate the molecule entirely.  Convincing evidence of SNR-MC interactions must come from other sources in these cases.

Internal flows within SNRs are expected to become highly turbulent where the SNR shock front interacts with the dense ISM \citep{jj99}.  Molecular lines toward SNRs, such as the rotational transitions of $^{12}$CO, should be velocity-broadened as a result of increased turbulence from a SNR-MC interaction.  Shocked MCs are also known to exhibit asymmetric line profiles where these interactions occur \citep{rr99,koo01,jcws10}.  As opposed to the presence of OH $1720~\text{MHz}$ maser emission, the detection of velocity-broadened, broad molecular line (BML) $^{12}$CO relies on more detailed spectroscopic information, especially in faint line ``wings'' where turbulent broadening is usually most apparent. At the same time, this type of observation offers insight into SNR evolution and the physical properties of both SN ejecta and MCs.

Even if the studies of OH maser emission, BML emission, and other shock indicators are all taken together, only about 60 Galactic SNR-MC interactions have been identified. However, to date there has been no large-scale systematic survey for shock-broadened molecular lines. Instead, the interactions have not been identified in a homogeneous manner, but are generally a result of investigations of individual SNRs, notably IC 443, W28, W44, CTB 109, Kes 69, Kes 79, W51C, and Cassiopeia A (Cas A) \citep{wootten77,den79b,tfis90,gd92,km97,atst99,rr99,zhou+09,kilpatrick+14}.  These sources span a wide range of ages and morphologies, from young, shell-like SNRs such as Cas A, through middle-aged and older ($> 10^{4}~\text{yr}$) remnants, and including mixed-morphology remnants and plerions.  A systematic and unbiased survey of broadened CO emission toward Galactic SNRs may yield a larger incidence of such interactions.

The work described in this paper uses CO emission for the first large-scale, systematic search for BML regions toward SNRs.  To interpret these observations, we developed an algorithm to identify regions of broadened $^{12}$CO $J=2-1$ emission where they exist toward each SNR-MC system.  We justify these identifications using the approximate systemic velocity at the distance inferred to each SNR, an estimate of the ``broadened'' CO emission, and the radio continuum morphology of the remnant itself.  In this way, we have identified BML regions toward nineteen SNRs in our sample, including nine newly identified candidate SNR-MC interactions.

\section{Source Selection and Distances}\label{sec:ssobs}

\begin{deluxetable*}
{lccccccccccc}
\tabletypesize{\scriptsize}
\tablecaption{Supernova Remnants in Survey\label{tab:survey}}
\tablewidth{0pt}
\tablehead{
Cat. No. & Name & $\alpha$ (J2000) & $\delta$ (J2000) & Region & RMS & Distance & BML? & Type & Pulsar & TeV $\gamma$-ray ($\sigma$) & Ref. \\
& & (h~m~s) & (d~m) & & (K) & (kpc) & & & &
}
\startdata
G4.5$+$6.8		& Kepler 	& $17~30~42$ & $-21~29$ & $10' \times 10'$ 		& 0.17 	& 4.0 			& 		& S 	& 	& HESS (0.68)	& 			\\ %
G08.3$-$0.0 	&			& $18~04~34$ & $-21~49$ & $10' \times 10'$ 		& 0.08 	& 16.3 			& D 	& S 	& 	& HESS (14.6)	& 			\\ 
G09.9$-$0.8 	&			& $18~10~41$ & $-20~43$ & $13' \times 10'$ 		& 0.11 	& 6.4 			& D 	& S 	& 	& HESS (-0.1)	& 			\\ 
G10.5$-$0.0 	&			& $18~09~08$ & $-19~47$ & $10' \times 10'$ 		& 0.15 	& 14.4 			& 	 	& S 	& 	& HESS (4.6)	& 			\\ 
G11.2$-$0.3 	&			& $18~11~27$ & $-19~25$ & $10' \times 10'$ 		& 0.09 	& 4.4 			& D 	& C 	& Y & HESS (5.5)	& 1			\\ 
G11.8$-$0.2 	&			& $18~12~25$ & $-18~44$ & $10' \times 10'$ 		& 0.09 	& 19.4 			& 		& S 	& 	& HESS (1.7)	&			\\ %
G12.0$-$0.1 	&			& $18~12~11$ & $-18~37$ & $7' \times 7'$ 		& 0.12 	& 10.9			&		& S		&	& HESS (0.7)	& 			\\ %
G12.2$+$0.3 	&			& $18~11~17$ & $-18~10$ & $10' \times 10'$ 		& 0.09 	& 15.6 			& D 	& S 	& 	& HESS (2.4)	& 			\\ 
G12.5$+$0.2 	&			& $18~12~14$ & $-17~55$ & $10' \times 10'$ 		& 0.12 	& 15 			& 		& C 	& 	& HESS (4.7)	& 			\\ 
G12.8$-$0.0 	&			& $18~13~37$ & $-17~49$ & $10' \times 10'$ 		& 0.11 	& 4.8			& 		& C 	& Y & HESS (24.3)	& 2			\\ 
G13.5$+$0.2 	&			& $18~14~14$ & $-17~12$ & $10' \times 10'$ 		& 0.12 	& 13.3 			& 		& S 	& 	& HESS (3.3)	& 			\\ 
G14.3$+$0.1 	&			& $18~15~58$ & $-16~27$ & $10' \times 10'$ 		& 0.11 	& 18.7 			& 		& S 	& 	& HESS (1.8)	& 			\\ %
G15.9$+$0.2 	&			& $18~18~52$ & $-15~02$ & $8' \times 7'$ 		& 0.11 	& 8.5 			& 		& S 	& 	& HESS (1.4)	& 			\\ %
G16.7$+$0.1 	&			& $18~20~56$ & $-14~20$ & $10' \times 10'$ 		& 0.15 	& 10		 	& DP 	& C 	& 	& HESS (4.7)	& 3,4,5		\\ 
G17.0$-$0.0 	&			& $18~21~57$ & $-14~08$ & $10' \times 10'$ 		& 0.18 	& 18.1 			& 		& S 	& 	& HESS (6.6) 	& 			\\ 
G18.1$-$0.1 	&			& $18~24~34$ & $-13~11$ & $8' \times 8'$ 		& 0.13 	& 5.6 			& 		& S 	& 	& HESS (6.9)	& 			\\ 
G18.6$-$0.2 	&			& $18~25~55$ & $-12~50$ & $6' \times 6'$ 		& 0.13 	& 13.2 			& D 	& S 	& 	& HESS (2.8)	& 			\\ 
G21.5$-$0.1 	&			& $18~30~50$ & $-10~09$ & $10' \times 10'$ 		& 0.09 	& 18.9 			& 		& S 	& 	& HESS (1.2)	& 			\\ %
G21.5$-$0.9 	&			& $18~33~33$ & $-10~35$ & $10' \times 10'$ 		& 0.07 	& 4.8 			& 		& C 	& Y & HESS (6.4)	& 7			\\ 
G23.6$+$0.3 	&			& $18~33~03$ & $-08~13$ & $8' \times 8'$ 		& 0.12 	& 6.9  			& D 	& ? 	& 	& HESS (1.5)	& 			\\ 
G27.4$+$0.0 	& 4C$-$04.71& $18~41~19$ & $-04~56$ & $10' \times 10'$ 		& 0.10 	& 8.5 			& D 	& S 	& Y & HESS (6.0)	& 3,8		\\ 
G29.6$+$0.1 	&			& $18~44~52$ & $-02~57$ & $8' \times 8'$ 		& 0.16 	& 10 			& D 	& S 	& Y & HESS (4.5)	& 9			\\ 
G29.7$-$0.3 	& Kes 75 	& $18~46~25$ & $-02~59$ & $10' \times 10'$ 		& 0.13 	& 6.0 			& DP 	& S 	& Y & HESS (10.1)	& 3,10,11	\\ 
G30.7$-$2.0 	&			& $18~54~25$ & $-02~54$ & $17' \times 16'$ 		& 0.14 	& 4.7 			& 		& ? 	& 	& HESS (0.9)	& 			\\ %
G31.5$-$0.6 	&			& $18~51~10$ & $-01~31$ & $19' \times 18'$ 		& 0.11 	& 6.3 			& 		& S 	& 	& HESS (1.9)	& 			\\ %
G31.9$+$0.0 	& 3C 391 	& $18~49~25$ & $-00~55$ & $10' \times 10'$ 		& 0.10 	& 7.2 			& DP 	& S 	& 	& HESS (0.0)	& 5,12,13,14\\ 
G32.4$+$0.1 	&			& $18~50~05$ & $-00~25$ & $10' \times 9'$ 		& 0.10 	& 17 			& D 	& S 	& 	& HESS (2.2)	& 			\\ 
G33.2$-$0.6 	&			& $18~53~50$ & $-00~02$ & $19' \times 18'$ 		& 0.11 	& 5.7 			& 		& S 	& 	& HESS (2.5)	& 			\\ 
G33.6$+$0.1 	& Kes 79	& $18~52~48$ & $+00~41$ & $10' \times 10'$ 		& 0.13 	& 7.1 			& DP 	& S 	& Y & HESS (2.7)	& 3,15,16	\\ 
G36.6$+$2.6 	& 			& $18~48~49$ & $+04~26$ & $19' \times 13'$ 		& 0.09 	& 8.7 			& 		& S 	& 	& HESS (-1.4)	& 			\\ %
G39.2$-$0.3 	& 3C 396 	& $19~04~08$ & $+05~28$ & $10' \times 10'$ 		& 0.11 	& 6.2		 	& DP 	& C 	& Y?& HESS (0.5)	& 3,17,18,19\\ 
G41.1$-$0.3 	& 3C 397 	& $19~07~34$ & $+07~08$ & $10' \times 10'$ 		& 0.10 	& 10.3			& DP 	& S 	& 	& HESS (3.0)	& 3,20		\\ 
G43.3$-$0.2 	& W49B 		& $19~11~08$ & $+09~06$ & $10' \times 10'$ 		& 0.09 	& 2.5			& DP 	& S 	& Y?& HESS (1.0)	& 3,21,22	\\ 
G54.1$+$0.3 	&			& $19~30~31$ & $+18~52$ & $10' \times 10'$ 		& 0.09 	& 6.2 			& P 	& C 	& Y & HESS (3.8)	& 3,17,23,24\\ 
G57.2$+$0.8 	& 4C-21.53 	& $19~34~59$ & $+21~57$ & $13' \times 10'$ 		& 0.19 	& 8.2		 	& 		& S 	&   & HESS (0.8)	& 			\\ %
G59.8$+$1.2 	&			& $19~38~55$ & $+24~19$ & $24' \times 10'$ 		& 0.11 	& 7.3		 	& 		& F 	& 	& HESS (1.6)	& 			\\ %
G63.7$+$1.1 	&			& $19~47~52$ & $+27~45$ & $10' \times 10'$ 		& 0.12 	& 3.8 			&   	& F 	& 	&			   	& 			\\ %
G69.0$+$2.7 	& CTB 80 	& $19~53~20$ & $+32~55$ & $10' \times 10'$ 		& 0.07 	& 1.5 			& 		& F 	& Y & MAGIC (0.6)	& 25		\\ 
G69.7$+$1.0 	&			& $20~02~40$ & $+32~43$ & $16' \times 19'$ 		& 0.14 	& 7.1 			& 		& S 	& 	&			   	&			\\ %
G74.9$+$1.2 	& CTB 87 	& $20~16~02$ & $+37~12$ & $8' \times 6'$ 		& 0.08 	& 6.1			& 		& F 	& Y?& VERITAS (6.2) & 17,21		\\ %
G76.9$+$1.0 	&			& $20~22~20$ & $+38~43$ & $10' \times 10'$ 		& 0.10 	& 10 			& 		& C 	& Y &				& 17		\\ %
G83.0$-$0.3 	&			& $20~46~55$ & $+42~52$ & $10' \times 10'$ 		& 0.15 	& 11.9 			& 		& S 	& 	&				&			\\ %
G84.2$-$0.8 	&			& $20~53~20$ & $+43~27$ & $22' \times 16'$ 		& 0.11 	& 6.0 			& 		& S 	& 	&				& 			\\ %
G111.7$-$2.1	& Cas A 	& $23~23~26$ & $+58~48$ & $10' \times 10'$ 		& 0.05 	& 3.4 			& DP 	& S 	& Y & MAGIC (5.2)	&26,27,28,29\\ 
G116.9$+$0.2	& CTB 1 	& $23~59~10$ & $+62~26$ & see \autoref{sec:obs} & 0.09 	& 3.1 			& 		& S 	& Y &				& 30		\\ 
G120.1$+$1.4 	& Tycho 	& $00~25~18$ & $+64~09$ & $10' \times 10'$ 		& 0.10 	& 2.5 			& 		& S 	& 	& VERITAS (5.8)	& 			\\ 
G130.7$+$3.1 	& 3C 58		& $02~05~41$ & $+64~49$ & $10' \times 10'$ 		& 0.08 	& 2.0 			& 		& F 	& Y & MAGIC (5.7)	& 31		\\ %
G132.7$+$1.3 	& HB 3 		& $02~17~40$ & $+62~45$ & see \autoref{sec:obs} & 0.11 	& 2.0		 	& DP 	& S 	& Y &				& 32,33		\\ 
G184.6$-$5.8 	& Crab		& $05~34~31$ & $+22~01$ & $10' \times 10'$ 		& 0.10 	& 2.0 			& 		& F 	& Y & HESS (129)	& 34		\\ 
G189.1$+$3.0 	& IC 443 	& $06~17~00$ & $+22~34$ & $9' \times 8'$ 		& 0.09 	& 1.5 			& DP 	& C 	& Y?& VERITAS (8.3)	& 35,36		 
\enddata
\tablecomments{Distances are those we adopt for this paper, as described in \autoref{sec:dist}.  For BML Detection, D: Detected broadened CO feature in this paper, P: Previous SNR-MC interactions as reported in \citet{jcws10}.    For Type, we report the SNR morphology from \citet{green14} where S: shell, C: composite, and F: filled except for G12.0$-$0.1 which is described as shell-like in \citet{kassim92}, G59.8$+$1.2 which is described as a plerion/filled shell SNR in \citet{sun11}, and CTB 80 which is described as an extended plerion/filled shell SNR in \citet{angerhofer+81}.  SNRs whose types have not yet been identified are indicated with a ?.  For pulsar, Y: SNR has a confirmed (i.e., timed) pulsar, Y?: SNR has a coincident, compact X-ray or radio source.  We include TeV gamma-ray detections by experiment and significance of detection in parentheses for HESS \citep{aharonian+06,aharonian+08,bochow11}, MAGIC \citep{albert+07a,aleksic+14}, and VERITAS \citep{acciari+09,acciari+11,aliu11}.  For references related to detections of pulsars/compact objects as well as previous evidence for SNR-MC interactions: (1) \citep{torii+97}; (2) \citep{gh09}; (3) \citep{green97}; (4) \citep{rm00}; (5) \citep{hyw08}; (7) \citep{gupta05}; (8) \citep{vg97}; (9) \citep{vgtg00}; (10) \citep{gvbt00}; (11) \citep{scyk09}; (12) \citep{rm93}; (13) \citep{frail96}; (14) \citep{rr99}; (15) \citep{gd92}; (16) \citep{ghs05}; (17) \citep{bl96}; (18) \citep{hrar09}; (19) \citep{scyk11}; (20) \citep{jcws10}; (21) \citep{grak96}; (22) \citep{krrj07}; (23) \citep{clbg02}; (24) \citep{lee+12}; (25) \citep{kulkarni+88}; (26) \citep{tananbaum99}; (27) \citep{ll99}; (28) \citep{hines04}; (29) \citep{kilpatrick+14}; (30) \citep{hc95}; (31) \citep{murray+02}; (32) \citep{lorimer+98}; (33) \citep{routledge91}; (34) \citep{lovelace+68}; (35) \citep{olbert+01}; (36) \citep{cornett77}}
\end{deluxetable*}

The SNRs in our survey were selected from those accessible with the Heinrich Hertz Submillimeter Telescope (SMT) on Mt. Graham, Arizona, $(\lambda, \phi) = (-109.8920, +32.7013)$.  In general, all SNRs with declination greater than $-22^{\circ}$ were considered, yielding a sample between Galactic longitude $+4.5$ and $+190$ degrees.  There are currently 160 known SNRs in this volume \citep{green14}.  From this larger sample, we selected SNRs with an apparent size less than $20'$ in diameter along their largest axis.  The targets we observed are listed in \autoref{tab:survey}.

Some targets listed in \autoref{tab:survey} were selected for comparison to our larger sample.  These SNRs are well-studied and large (diameter $> 20'$) and in these cases we mapped a smaller subregion of the entire remnant.  These targets include CTB 80, CTB 1, HB 3, and IC 443.  IC 443 is a well-known SNR-MC system, while HB 3 is suspected to have SNR-MC interactions \citep{routledge91}.

\subsection{Distances}\label{sec:dist}

Reliable distances are critical when characterizing SNRs.  In order to draw an association between broad-line MCs and the SNRs themselves, a variety of distance indicators exist such as proper motion estimates, radial velocity, \HI\ absorption, and OB associations, each with varying degrees of uncertainty.  Where the properties of SNRs and their environments differ, distance indicators can yield estimates with significant disagreement.  For example, the distance to G4.5$+$6.8 (Kepler/SN 1604) has historically been controversial, with estimates ranging from around $4 - 6$ kpc \citep{ld83, wl83, rg99, vink08, csv12} to $10 - 12$ kpc \citep{milne70, il72, bbs92}, although proper motion estimates now favor the closer distance.  This lack of agreement between various methods underscores the systematic uncertainties present in all distance indicators, from discrepant optical absorption \citep{dg80} to disagreement in the maximum optical brightnesses of historical SNe \citep{vmt73}.

For these reasons, we carefully chose distance indicators from among the literature, favoring those determined from proper motion estimates, kinematic estimates from \HI\ absorption and CO associations, and line-of-sight X-ray absorption.  The values we chose are given in \autoref{tab:survey} and described in \autoref{sec:altdist}.  However, for nineteen of the remnants in our survey, we were unable to find reliable distance indicators in the literature.  For these objects, we resort to the surface brightness-size or $\Sigma - D$ relation \citep{shklovskii60, cc76}.  This empirical relationship describes SNR evolution with time, whereby the radio surface brightness ($\Sigma$) decreases monotonically while the physical diameter ($D$) increases monotonically.  Thus, a power law relationship can be fit such that

\begin{equation}
\Sigma = A D^{\beta}
\end{equation}

\noindent $A$ and $\beta$ are calibrated using SNRs whose distances are known from other methods.  Measuring the radio surface brightness and angular size of the SNR then yields an angular diameter distance.  \citet{pdvu14} fit this relationship and derive distances to 225 Galactic SNRs.  The authors find the best-fit slope to be $\beta \approx -5.2$.  This method involves significant systematic uncertainties and should be used with caution.  Distance measurements for some of the SNRs in our sample are taken from this study where we adopt the authors' most probable distance (i.e., the orthogonally fit value from Table 4 in \citet{pdvu14}).  These include sixteen out of the 50 SNRs we observed.  We provide our own estimates for the remaining three SNRs, which we describe in \autoref{sec:altdist}.

\section{Observations}\label{sec:obs}

\begin{figure*}
	\centering
	\includegraphics[width=\textwidth]{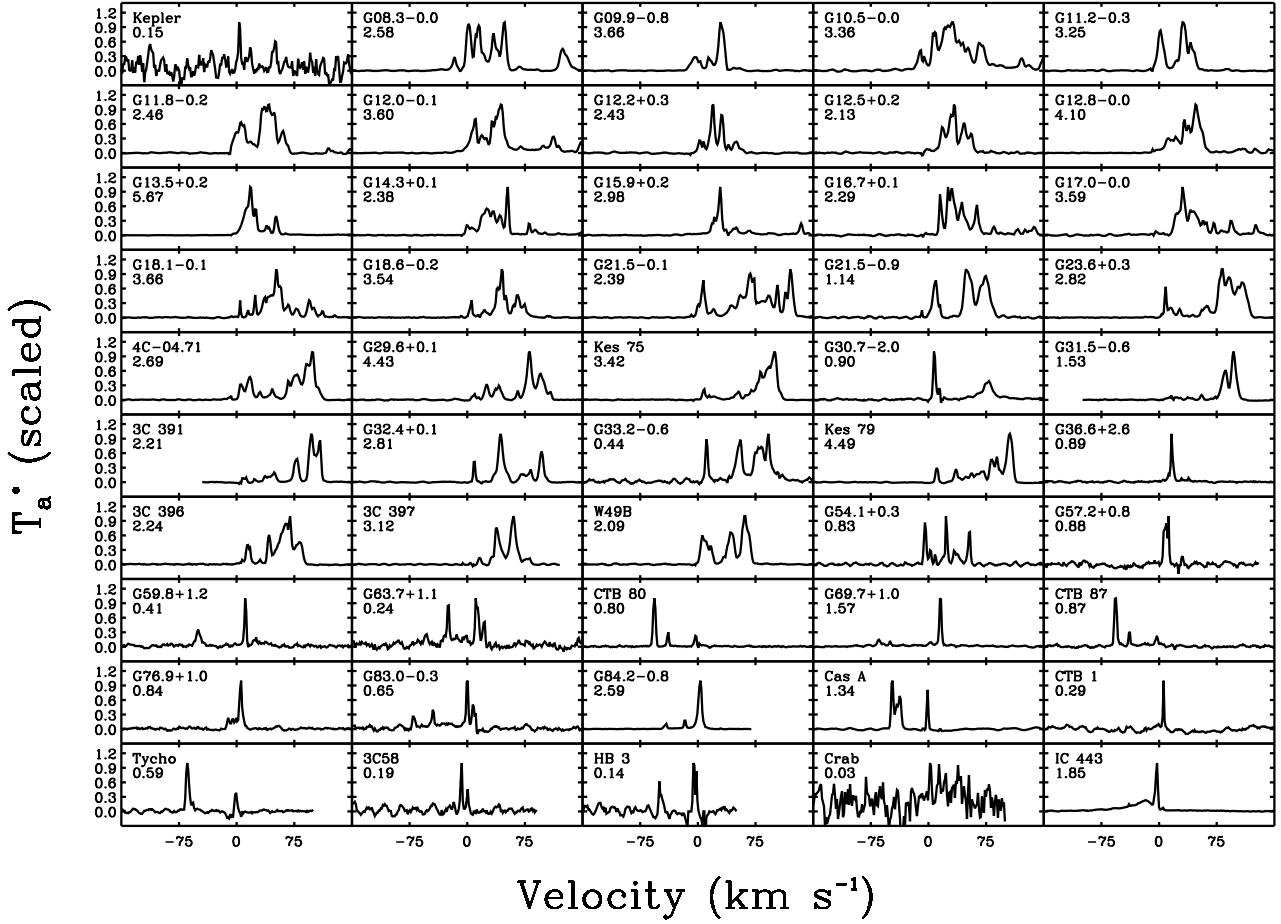}
	\caption{\scriptsize $^{12}$CO $J=2-1$ emission toward all 50 SNRs in our sample averaged over each OTF map from $-150$ to $+150~\kms$.  The spectra are scaled such that the maximum antenna temperature in each spectrum has $\text{T}_{a}^{*} = 1$.  In the upper left of each panel, we give the name of the remnant and the scale factor where $\text{T}_{a}^{*} (\text{K}) = \text{T}_{a}^{*} (\text{scaled}) \times \text{scale factor}$.
}\label{fig:all}
\end{figure*}

We mapped the vicinity of 50 SNRs (\autoref{tab:survey}, \autoref{fig:all}) with the 10m SMT using the ALMA-type sideband separating 1.3 mm receiver between 2013 April 21 and 2015 May 8.  Apart from the size and average RMS noise of each map as given in \autoref{tab:survey}, the observations and reduction for each remnant were identical to those described in \citet{kilpatrick+14}.

As discussed in \autoref{sec:ssobs}, there were several targets for which we did not observe the entire remnant.  Rather, we targeted either a region we knew to have broadened lines or an X-ray (i.e., $0.5 - 10~\text{keV}$) bright region toward any known molecular emission.  These objects were used either for comparison to other targets (HB 3, IC 443) or as a targeted search for new BML regions toward large remnants (CTB 80, CTB 1).  In particular, we targeted two regions of X-ray bright emission from CTB 1, centered at $\hms{23}{59}{05},~\dms{+62}{30}{00}$ with dimensions ($\alpha \times \delta$) $10' \times 30'$ and at $\hms{0}{03}{05},~\dms{+62}{45}{00}$ with dimensions $10' \times 20'$.  Similarly, for HB 3 we targeted two regions centered at $\hms{2}{16}{15},~\dms{+63}{40}{00}$ with dimensions $10' \times 30'$ and at $\hms{2}{19}{15},~\dms{+63}{50}{00}$ with dimensions $10' \times 10'$.  These regions were centered on local X-ray enhancements as reported by \citet{lazendic+06}.  For the other large targets, we used single fields as reported in \autoref{tab:survey}.  We analyzed these four remnants and we indicate whether BML regions were detected in each case.  However, given the fact that we did not map the entirety of each remnant and some were selected on the basis of previous evidence for SNR-MC interactions, we discount all four in our discussion of SNR-MC interactions (except in \autoref{sec:gamma}) as a possible source of bias.

\section{Algorithm for Selecting BML Regions from Observations}\label{sec:algorithm}

\subsection{Data}\label{sec:aldata}

In our analysis, we used the $1~\text{MHz}$ resolution data, which has 512 channels for a velocity range of $667~\kms$ at $230~\text{GHz}$.  For a receiver tuned to the $^{12}$CO $J=2-1$ and centered at a systemic velocity of $0$, we can be reasonably certain that all Galactic molecular emission along the line-of-sight is observed.

\begin{figure*}
	\begin{center}
		\begin{minipage}[t][][b]{0.42\textwidth}
			\includegraphics[width=\textwidth,angle=270,origin=c]{3C391_region03.eps}
		\end{minipage}
		\hfill
		\begin{minipage}[t][][b]{0.42\textwidth}
			\includegraphics[width=\textwidth]{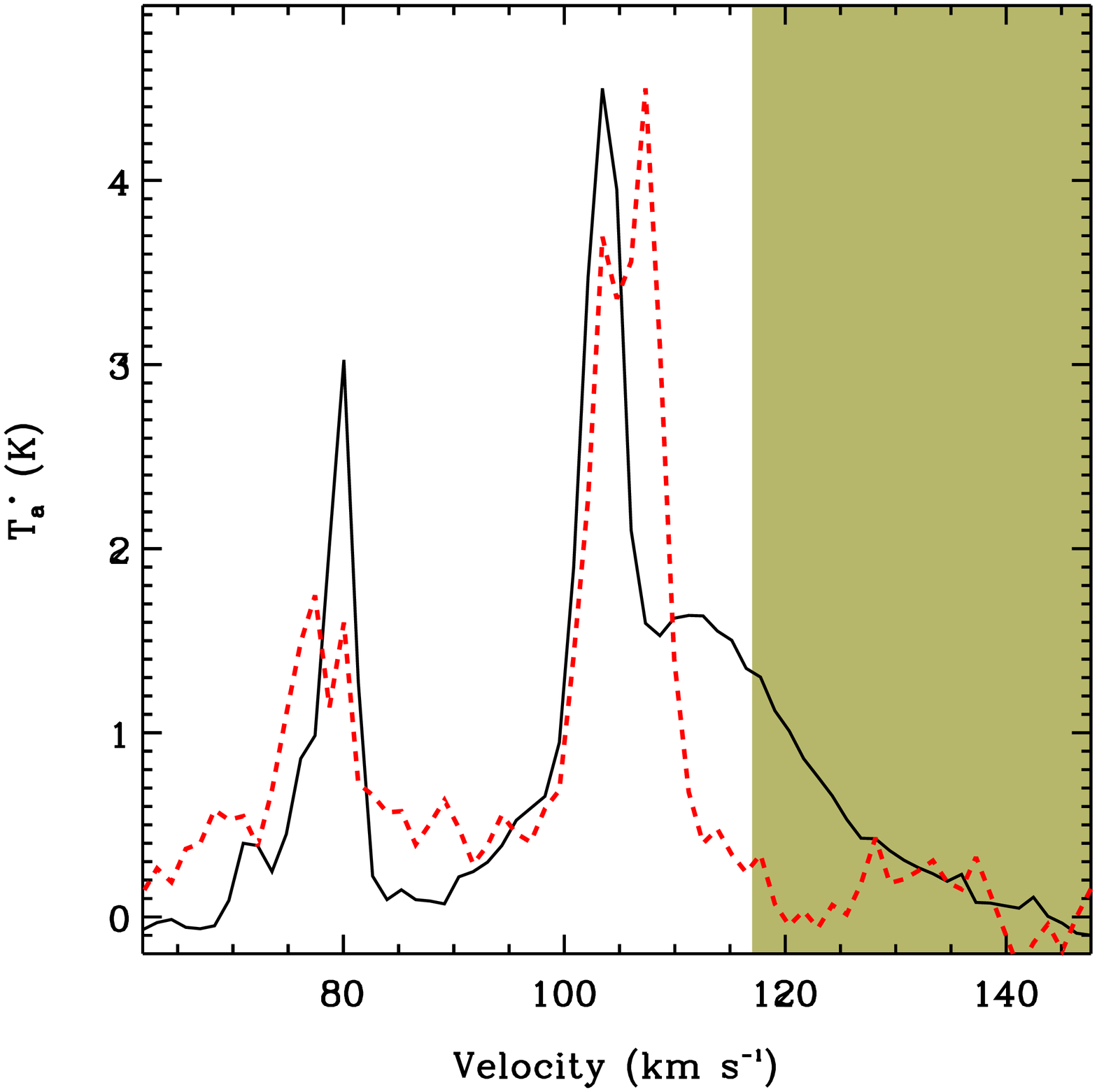}
		\end{minipage}
	\end{center}
	\vspace{-0.5in}
		\caption{\scriptsize (Left) Integrated $^{12}$CO $J=2-1$ emission toward 3C 391 from $+102$ to $+108~\kms$.  Orange contours are 20 cm radio continuum emission of the remnant from \citet{rm93}.  Pixels with broadened molecular emission are shown as white squares.  We denote a comparison region of unbroadened, bright molecular emission with a larger red square.  (Right) The average spectrum from these pixels at the velocity of the BML identification (solid black) along with a spectrum of the unbroadened, bright comparison region averaged over the pixels denoted by the red square (dashed red).  Shaded parts of the spectrum correspond to velocities ruled out by the inferred distance to the SNR.}\label{fig:3c391}
\end{figure*}

The expected velocity range of the CO spectrum at each distance is determined by assuming a $50\%$ uncertainty on the distances we assume to each SNR, which agrees with typical fractional errors for distances calculated using the $\Sigma-D$ relation \citep{pdvu14}.  Comparison with the Clemens rotation curve \citep{clemens85} allows us to limit the velocity range over which MCs might exist that are interacting with a SNR.  We extend this range by $15~\kms$ at either end of the range in order to account for systematic errors in finding BML regions associated with SNRs (see, e.g., \autoref{sec:3c391}).  \autoref{fig:3c391} demonstrates this method for the known BML region toward 3C 391.  The distance to this SNR is taken to be $7.2 \pm 3.6$ kpc, which at Galactic coordinates $l = 31.9, b = 0.0$ corresponds to a systemic velocity range of $+59$ to $+102~\kms$ or $+44$ to $+117~\kms$ with the added velocity range.

\subsection{Summary of BML Region Algorithm}\label{sec:alsum}

Within the range of permitted velocities for each SNR, we imposed criteria on the line width and profile of observed molecular emission which we describe and justify in \autoref{sec:alpos}.  In summary, these criteria are: 

\begin{itemize}
\item We required CO emission to be detected at $3~\text{T}_{rms}$ (the root mean square of emission in the line-free spectrum at each pixel as in \autoref{tab:survey}) in every spectral channel we used to calculate the line profile.
\item We used only five spectral channels for the width determination.
\item We required that the line profile not include secondary maxima.
\item We required the FWHM of observed $^{12}$CO lines $> 6~\kms$.
\end{itemize}

The purpose of these criteria was to identify regions of molecular emission that correspond with disturbed molecular gas where a SNR-MC interaction has occurred.  At the same time, we attempted to minimize false positives, e.g. due to molecular emission overlapping in velocity-space, and false negatives, e.g. where interactions are subtle and exhibit small velocity-widths, the CO line is weak, or strongly asymmetric line profiles are present.

\subsection{Criteria for Possible Interactions}\label{sec:alpos}

For clouds with sizes on the order of a few parsecs, quiescent $^{12}$CO line widths are typically $\sim 2 - 3~\kms$ \citep{hb04}.  Therefore, we adopted a threshold of a $6~\kms$ FWHM to indicate BML regions.  We evaluate and discuss the validity of this threshold in \autoref{sec:casa}.

Furthermore, BML regions can have asymmetric line profiles where the molecular gas has been shocked and accelerated.  In addition to large line widths (i.e., $> 10~\kms$) at negative or positive velocities relative to the peak, the line profile is characterized by faint wings where turbulent acceleration of molecular gas is apparent.  This line profile can be confused with multiple MCs along the same line-of-sight (e.g., toward the Galactic Center) resulting in superimposed emission in the spectrum of a small region or even a single pixel.  Thus, identifying BML regions ideally relies on as small a region as possible with as few spectral channels as possible.

To account for line widths, asymmetry, and molecular emission from multiple sources along a line-of-sight, we developed an algorithm for identifying BML regions in $^{12}$CO $J=2-1$.  Initially, we constructed a list of individual spectral channels by examining the spectrum at each pixel.  Where the $^{12}$CO emission in individual spectral channels exceeded $10~\text{T}_{rms}$ and represented a local maximum, we flagged that channel for analysis.  These flagged spectral features span the entire data cube and represent the line peaks of individual emission features in our data.

At $230~\text{GHz}$ and for a spectrum with $1~\text{MHz}$ resolution, the $6~\kms$ velocity-width threshold we adopted corresponds to a feature whose FWHM is $4.6~\text{MHz}$ or roughly five spectral channels.  Thus, to measure the line width of a broad feature, we required at least five channels or four channels in addition to the flagged local maximum of a spectral feature.  

Confusion where a single spectral channel contains emission from multiple MCs can be problematic.  In part, we minimized confusion by limiting the number of channels we used to measure line widths to the absolute minimum.  However, this method does not preclude the scenario in which two molecular features with narrow widths (i.e., $> 3~\kms$) are separated by less than their half widths in velocity-space.  Therefore, in addition to the requirements described above, our algorithm stipulates that the emission on either side of the line peak must decrease monotonically in each channel and that the emission in each channel exceeds $3~\text{T}_{rms}$.

Finally, using the five spectral channels around each flagged line peak, we measured the line widths using a method that accounts for possible asymmetries in the line profile.  We took the spectrum at a particular pixel $(x_{0}, y_{0})$ and channel $v_{0}$ corresponding to one of the flagged local maxima in our data cube. Here, we use the antenna temperature at a particular pixel, $\text{T}_{a}^{*}(v) = \text{T}_{a}^{*}(x_{0}, y_{0}, v)$. Next, we measured the velocity centroid $v_{c}$ of the line peak around $v_{0}$ by calculating the first moment of the spectrum

\begin{equation}
v_{c} = \frac{\sum\limits_{i=-2}^2 v_{i} \text{T}_{a}^{*}(v_{i})}{\sum\limits_{i=-2}^2 \text{T}_{a}^{*}(v_{i})}
\end{equation}

\noindent We then evaluated the second moment of the spectrum in this range,

\begin{equation}
\sigma = \sqrt{\frac{\sum\limits_{i=-2}^2 \text{T}_{a}^{*}(v_{i}) (v_{i} - v_{c})^{2}}{\sum\limits_{i=-2}^2 \text{T}_{a}^{*}(v_{i})}}
\end{equation}

\noindent This statistic is the velocity dispersion of the spectrum around the line peak.  For a Gaussian line profile, the FWHM $= 2.355 \sigma$.  For comparison to other line indicators, we used the velocity-width statistic $\Delta v = 2.355 \sigma$.  If this value exceeds $6~\kms$, the peak of the corresponding line profile is flagged in our data cube as a BML.  For SNRs with newly identified BML regions, we give the total number of pixels with BML identifications in parentheses in \autoref{tab:new}.

\begin{deluxetable}
{lcccc}
\tabletypesize{\scriptsize}
\tablecaption{Newly Detected BML Regions Toward SNRs\label{tab:new}}
\tablewidth{0pt}
\tablehead{
Cat. No. & \# pix. (total) & $\bar{v_{c}}$ & $\Delta$v & $d_{v_{c}}$ ($d_{\text{Tab. 1}}$)\\
& & (km s$^{-1}$) & (km s$^{-1}$) & (kpc)}
\startdata
G08.3$-$0.0 &	7 (10)		& +2.6		& 7.3 & 16.4 (16.3)\\
G09.9$-$0.8 &	8 (10)		& +31		& 7.1 & 4.0 (6.4) \\
G11.2$-$0.3 &	15 (21)		& +32		& 8.1 & 7.2 (4.4) \\
G12.2$+$0.3 &	2 (3) 		& +50		& 6.9 & 11.8 (15.6)\\
G18.6$-$0.2 &	12 (15) 	& +42		& 8.5 & 12.7 (13.2)\\
G23.6$+$0.3 &	7 (11)		& +91		& 6.7 & 5.6 (6.9)\\
G27.4$+$0.0 &	12 (20)		& +100		& 9.0 & 9.0 (12.8)\\
G29.6$+$0.1 &	5 (5)		& +94		& 10.0& 9.0 (10)\\
G32.4$+$0.1 &	3 (4)		& +43		& 7.5 & 11.8 (18.5)
\enddata
\tablecomments{The number of pixels corresponds to the total number of pixels associated with BML regions (\autoref{sec:sum}, \autoref{sec:new}).  In parentheses, we report the total number of BML detections (\autoref{sec:alpos}), including those ruled out as potential BML regions as described in \autoref{sec:sum}.  The velocity estimate $\bar{v_{c}}$ is a signal-weighted averaged of the velocity centroid from all BML detections toward each SNR (\autoref{sec:sum}).  The velocity-width $\Delta v$ is an average of each BML detected toward the SNR.  The distances $d_{v_{c}}$ and $d_{\text{\autoref{tab:survey}}}$ are the distance to each SNR as determined from the rotation curve at $\bar{v_{c}}$ and the distance reported in \autoref{tab:survey}, respectively.  When calculating the former value, we prefer distances that are closer to the latter in order to resolve the distance ambiguity.}
\end{deluxetable}

\subsection{BML Region Identification}\label{sec:sum}

As described in \autoref{sec:alpos}, local maxima in the spectrum at each pixel are flagged if the line profile calculated at that point matches the outlined criteria.  In this way, we can construct a separate ``flagged'' data cube with dimensions equal to the original data cube, that is, the antenna temperature $\text{T}_{a}^{*}(x,y,v)$.  This ``flagged'' data cube has $F(x,y,v) = 1$ if the spectrum at position $(x,y)$ and in the velocity channel $v$ meets our criteria, and $F(x,y,v) = 0$ otherwise.

We can further break down the ``flagged'' data cube into regions by using a group-finding algorithm simultaneously in spatial coordinates and velocity-space.  For the purposes of this analysis, the definition of BML regions will ultimately depend on the physical significance of these groups.

As described in \autoref{sec:obs}, our $^{12}$CO $J=2-1$ observations were performed with a 10m diameter telescope at $230~\text{GHz}$ with a FWHM beam of $33''$.  Each data cube was constructed by taking observations spaced $10''$ apart.  In our analysis, detection of pixels with broad lines whose centroids lie within each others' FWHM and within a radius of 2 pixels represents a single BML region.  For BML regions that appear extended on the sky, we can define groups that span several pixels, provided the broad-line detection is approximately at the same velocity (i.e., within a window spanning $\Delta v$ measured for each BML).       

A velocity centroid is measured for each BML detection representing a local maximum in the $^{12}$CO spectrum.  However, each of these points is biased by the way in which we construct the original data cube $T_{a}^{*}$, and the determination of a group of BML detections must be physically significant.  In the same way, the velocity centroid of each local maximum must be tied in some way to the actual velocity centroid for the entire BML region with which it is associated.  Therefore, we utilized a signal-weighted average to measure this value, for which the BML region velocity centroid ($\bar{v_{c}}$) is defined by the number of local maxima it includes ($N$), the Gaussian amplitude of each local maximum ($T_{a,i}^{*}$), and the velocity centroid calculated ($v_{c,i}$ as described in \autoref{sec:alpos})

\begin{equation}
\bar{v_{c}} = \frac{\sum\limits_{i=1}^{N} T_{a,i}^{*} v_{c,i}}{\sum\limits_{i=1}^N T_{a,i}^{*}}
\end{equation}

For newly discovered BML regions toward SNRs, we report in \autoref{tab:new} the number of BML pixels we detect toward each remnant, the signal-weighted velocity centroid as described above, and the average velocity-width of all BML detections toward the SNR.  Finally, we recalculated the distance to each remnant using the signal-weighted velocity centroid as a prior for the Clemens rotation curve and we report this distance in \autoref{tab:new}.  In order to resolve the distance ambiguity, we chose the distance closest to our original estimate from \autoref{tab:survey}.

The agreement between the distances reported in \autoref{tab:survey} is variable, although nearly always within the $50\%$ uncertainty on the original estimate given the prior we used to find associated CO emission (\autoref{sec:aldata}).  In some cases, these values are very close to the original distances, especially for G08.3$-$0.0, G18.6$-$0.2, 4C$-$04.71, and G29.6$+$0.1, where the distances were estimated from the $\Sigma-D$ relation for the first three objects and X-ray \HI\ column density for the latter.

\subsection{BML Region Examples: 3C 391 and Cas A}\label{sec:applic}

We applied our algorithm to two test cases involving known SNR-MC systems: 3C 391 (\autoref{sec:3c391}) and Cas A (\autoref{sec:casa}).  In this way, we demonstrate the types of BML regions selected by our algorithm and the accuracy of the measurements.  These two systems present unique problems that must be addressed in identifying BML regions in a large sample of $^{12}$CO $J=2-1$ data.  Here, we discuss previously detected signatures of BML regions toward each remnant and the underlying SNR-MC interactions they may trace.  From these examples, we examine the extent to which our algorithm can be optimized.

\subsubsection{3C 391}\label{sec:3c391}

Several observations of $^{12}$CO targeted on 3C 391 have revealed the presence of broadened molecular features from $+90$ to $+110~\kms$ \citep{sanders86,wilner98,rr99}.  The emission exhibits significant broadening relative to the ambient gas, with $^{12}$CO $J=2-1$ line widths exceeding $20~\kms$ in some regions (\autoref{fig:3c391}).  This emission has consistently been interpreted as a signal of postshock gas.  

The molecular line profiles also exhibit a bright, narrow component ($\Delta v \sim 2~\kms$) at the same velocities and peaking in the same region.  This observation suggests an association between molecules of significantly different temperatures and densities and presents a challenge when identifying regions of shocked molecular gas.  One hypothesis for the presence of the narrow component is that it arises from cooler, less turbulent gas reforming behind a dissociative shock \citep{rr99}.  That is, while the narrow component may be associated with the SNR shock, in some regions it could be superimposed upon and dominate the line profile from the broadened emission.  Under this hypothesis, we might only detect BML regions where the SNR-MC interaction is ongoing or recent.  Thus, detection of recently shocked gas in $^{12}$CO may be extremely sensitive to the criteria for identifying individual line components (i.e., local maxima) in the spectrum.  In our analysis, we selected all the local maxima in velocity-space whose signal exceeds $10~\text{T}_{rms}$.  In this way, our criteria are unbiased by the coincidence of narrow and bright line components with broad and faint components as in 3C 391.  

One caveat is that false detections can occur where the local maxima are close together in velocity-space; for example, we might report no detection if the narrow component was within $2$ spectral channels of the broad component.  In this case, additional information on the line profile would be necessary to calculate an accurate line width.  While this scenario does not occur in 3C 391, one could imagine dissociated molecular gas reforming behind a shock whose narrow emission peaks at roughly the same systemic velocity as the undissociated, turbulently broadened component.  As a follow-up to our initial analysis, we modified our requirement for the five channels to include scenarios where the line profile does not decrease monotonically for two channels on either side of the local maxima.  While we still required five velocity channels, we extended the channel window on either side of the line profile to avoid contamination from other emitting regions in velocity-space.

As we demonstrate in \autoref{fig:3c391} (left), the well-studied, shocked MC toward 3C 391 is located to the south of the remnant along a region of bright radio continuum.  The $^{12}$CO spectrum in this region is significantly broadened in a feature centered around a systemic velocity of $+112~\kms$ as seen in \autoref{fig:3c391} (right).  Our data is of sufficient spatial and spectral resolution to resolve and confirm the narrow and broad spectral features toward this cloud.

Another important point that arises when examining similar BML regions is that the $+112~\kms$ line appears to lie outside the velocity range inferred for an object at $l = +31.9$ and $d = 7.2 \pm 3.6~\kms$, and the broad profile itself is at least $15~\kms$ from the maximum velocity allowed within this range.  The remnant appears to be near the tangent point of the Galactic rotation curve at this velocity ($+102~\kms$), but the MC is slightly redshifted with respect to this velocity.  Given that the shock interaction with the MC has been confirmed in multiple wavelengths and is generally assumed to be associated with the remnant, one hypothesis for the velocity discrepancy is that remnants may accelerate MCs in an ongoing interaction.  This point underscores the need for care in treating SNR-MC interactions whose distances are near the tangent point of the rotation curve along a particular line-of-sight.  As we describe in \autoref{sec:aldata}, we extend the velocity window used as a prior in our CO spectra by $15~\kms$ at either end to account for this potential problem toward other SNRs.

\subsubsection{Cas A}\label{sec:casa}

Cas A exhibits subtle indications of a SNR-MC interaction in several ways.  Warm CO gas (T $\sim 20~\text{K}$) directly to the west of the SNR implies some source of energy to heat the cloud \citep{wilson93}.  There is also a correlation between the brightest portions of the radio shell at $1~\text{GHz}$, nonthermal X-ray emission, and the radio spectral indices in the same region suggesting that the ejecta are encountering a denser medium \citep{ar96,kra96}.  In the near-infrared K$_{s}$ band, strong nonthermal emission occurs in the same regions \citep{rho03}.  These observations are consistent with an interaction scenario, although they do not conclusively point to the presence of a SNR-MC interaction.

Recent results indicate that hard X-ray emission above $15~\text{keV}$ toward Cas A is dominated by the western half of the remnant \citep{grefenstette15}.  Knots of bright emission in nonthermal emission are consistent with steep spectral indices in the X-ray and some change in the physical environment.  These knots are coincident with bright radio emission and the large MC to the west of Cas A.  Further evidence of an ongoing interaction scenario may come from changes in this and other nonthermal emission on short timescales.

In the MC itself, H$_{2}$CO absorption studies suggest that certain species are broadened to as much as $5.5~\kms$ \citep{rg02}.  Similarly, slight broadening in $^{12}$CO $J=1-0$ up to $4.5~\kms$ is observed \citep{ll99}, while line widths in $^{12}$CO $J=2-1$ range from $6.2 - 6.7~\kms$ \citep{kilpatrick+14}.  Thus, evidence for a SNR-MC interaction toward Cas A is strongly correlated in the X-ray, radio, near-infrared, and sub-millimeter, but the overall result in $^{12}$CO $J=2-1$ is subtle.  In order to detect BML regions toward similar SNR-MC interactions, we must rely on a precise constraint in the velocity-width.

\begin{figure*}
	\begin{center}
		\begin{minipage}[t][][b]{0.42\textwidth}
			\centering
			\includegraphics[width=\textwidth,angle=270,origin=c]{CasA_region01.eps}
		\end{minipage}
		\hfill
		\begin{minipage}[t][][b]{0.42\textwidth}
			\centering
			\includegraphics[width=\textwidth]{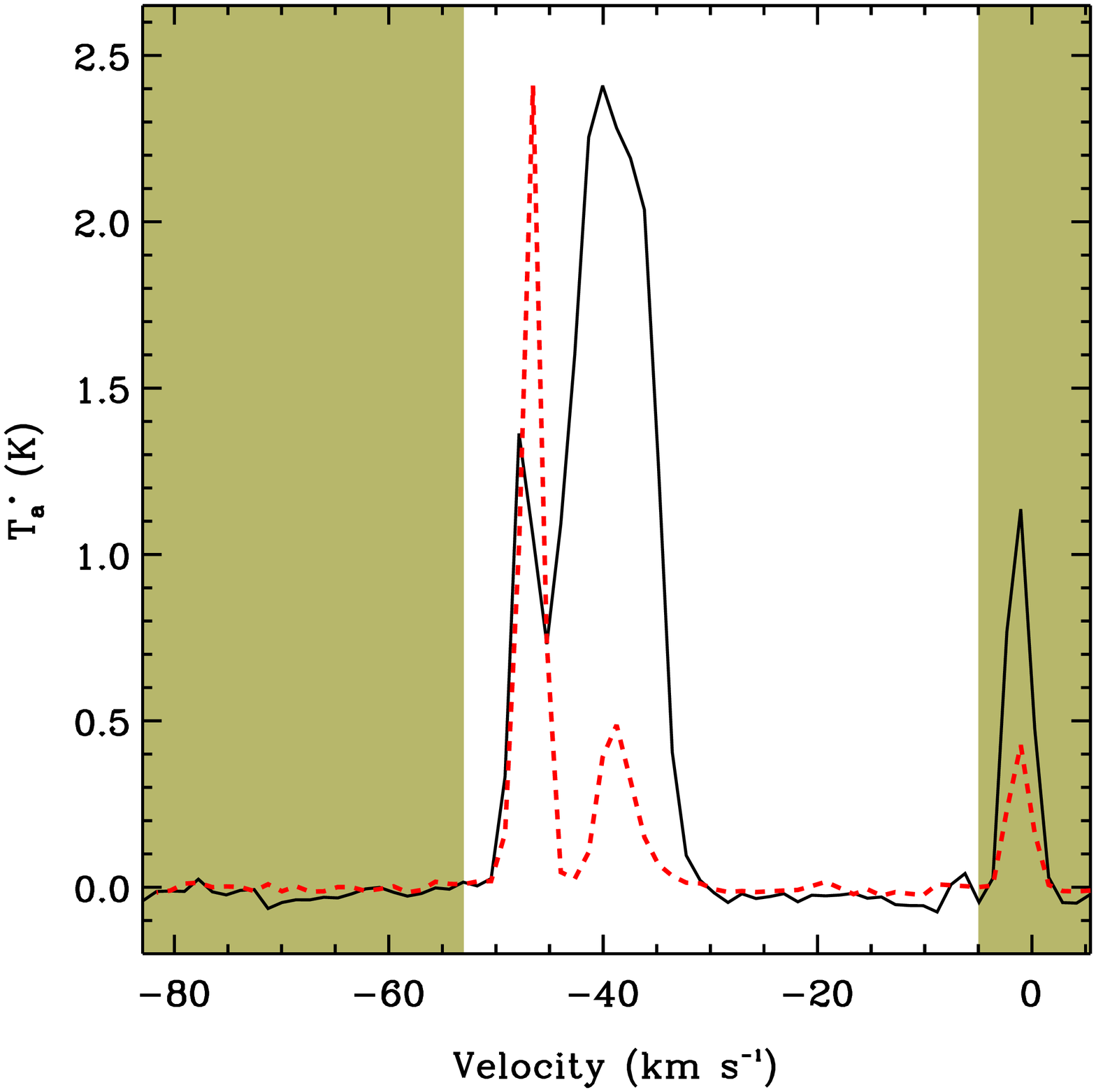}
		\end{minipage}
	\end{center}
	\vspace{-0.5in}
		\caption{\scriptsize (Left) Integrated $^{12}$CO $J=2-1$ emission toward Cas A from $-43$ to $-35~\kms$.  Orange contours are 20 cm radio continuum emission of the remnant from \citet{arlpb91}.  Pixels with broadened molecular emission are shown as white squares.  We denote a comparison region of unbroadened, bright molecular emission with a larger red square.  (Right) The average spectrum from these pixels at the velocity of the BML identification (solid black) along with a spectrum of the unbroadened, bright comparison region averaged over the pixels denoted by the red square (dashed red).  Shaded parts of the spectrum correspond to velocities ruled out by the inferred distance to the SNR.}\label{fig:casa}
\end{figure*}

Optimizing this constraint might be achieved by comparison with narrow molecular emission in the same field.  Line widths of narrow components in the Cas A field range from $2.5 - 3.4~\kms$, or about half of the ``broadened'' line widths.  Therefore, one criterion for ``broadened'' emission might be: can we divide these emission lines into distinct ``narrow'' and ``broad'' categories?  For example, how sensitive is the identification of BML regions to the velocity-width constraint we applied in \autoref{sec:alpos}?  We illustrate this point in \autoref{fig:casa} (right) where we compare broadened features to unbroadened emission as determined by our analysis.  The regions with the broadest $^{12}$CO line widths are qualitatively different from regions with narrow molecular lines.

Furthermore, we report the total number of pixels selected by the BML algorithm for newly identified BML regions in \autoref{tab:new}, which includes pixels that were ruled out as potential BML regions by the criteria described in \autoref{sec:sum}.  For newly identified BML regions, the algorithm selects these pixels with little contamination from other positions.  That is, the total number of BML pixels identified using the criteria described in \autoref{sec:alpos} is generally $75 - 85\%$ the number of BML pixels we determine are associated with the SNR as described in \autoref{sec:sum}.  We interpret this result to mean that a $6~\kms$ velocity-width threshold may be close to optimal for identifying BML regions toward SNRs and therefore minimizes false positives from spurious detections of, for example, overlapping molecular lines.

\subsection{Comparison to Other BML Region Indicators}

\citet{jcws10} compare molecular line studies to other traditional detection methods for SNR-MC interactions.  These include detection of the OH $1720~\text{MHz}$ maser, morphological correspondence between a SNR and a MC, molecular line emission with a large high-to-low line ratio such as $^{12}$CO $J=2-1/J=1-0$, detection of near-infrared emission from shock excitation such as [Fe II] or H$_{2}$ emission, and infrared colors consistent with molecular shocks.

It is valuable to utilize as many of these indicators as possible, and in our analysis we rely on morphological correspondence between the SNR in radio emission and the MC as well as reports of OH $1720~\text{MHz}$ maser emission.  However, the detection of broadened $^{12}$CO line emission is a unique tool for tracing SNR-MC interactions in our galaxy.  Spectroscopic information is vital for identifying the systemic velocity of molecular gas and therefore associating the kinematic distance to molecular emission with distances to SNRs.  This information reduces the probability of confusion by any chance alignment of a shock tracer with a SNR.  Carbon monoxide is also the second most abundant molecule and possesses one of the highest bond-dissociation energies of any diatomic molecule.  Therefore, we would expect CO to be more easily visible and less susceptible to dissociation behind a SNR shock than OH, H$_{2}$ or any other molecular shock tracer.  Finally, broad, low-J CO emission is a direct tracer of disturbed gas in a MC, as opposed to line emission from atomic species or near-infrared colors which do not directly or unambiguously trace the molecular gas.  Probing the MC for molecular shocks and then drawing an association with the SNR is well-suited for finding SNR-MC interactions, even in a large survey. 

\section{Results}

We applied the BML region algorithm to the $^{12}$CO $J=2-1$ spectra (\autoref{fig:all}) toward all 50 SNRs in our sample to determine the location of broadened CO features.  The algorithm revealed broad-line detections toward nineteen SNRs, including nine previously unidentified SNR-MC systems.  For the other ten systems with coincident BML regions, we discuss our detections toward known SNR-MC systems in the context of previous studies in \autoref{sec:known}.  One of the remaining 31 SNRs, G54.1$+$0.3, had previous indications of a BML region.  We argue in \autoref{sec:non} that the associated MC may not be velocity-broadened.  Finally, in \autoref{sec:new}, we discuss the results of our algorithm for newly identified SNR-MC systems and compare these detections to the observed properties of the SNRs themselves.  For reference, we include a discussion of previous observations for some remnants in \autoref{sec:anc}, where we discuss details of the radio continuum, OH maser detections, and pulsar observations.

\subsection{SNRs with Previous Detections Supporting SNR-MC Interactions}\label{sec:known}

BML regions were detected toward ten SNRs with known SNR-MC systems.  Here we compare the characteristics of the detected BML regions to results from previous studies.

\textit{G16.7$+$0.1} was observed to have a SNR-MC interaction by \citet{rm00}.  The authors report detection of a small MC in $^{12}$CO $J=1-0$ at $+25~\kms$ and coincident with OH maser detections to the south of the remnant.  The CO emission from this cloud is seen from $+25.1$ to $+25.9~\kms$.  The cloud is small ($2.1 \times 10^{3}$ M$_{\odot}$, n$_{\text{H}_{2}} = 260$ cm$^{-3}$) and has a CO line width of only $1.5~\kms$, consistent with the line width of an ambient MC approximately $3$ pc in size \citep{hb04}.

\citet{rm00} report detection of two other MCs toward G16.7$+$0.1, which they call the ``central'' and ``northwestern'' clouds, at systemic velocities between $+25$ and $+27~\kms$.  The central cloud is located directly to the east of the brightest $5~\text{GHz}$ enhancement in the remnant, implying a possible connection between nonthermal emission and the MC.  The reported morphological correspondence, small size of the central cloud ($1.8 \times 10^{3}$ M$_{\odot}$), and moderately broadened CO profile ($\Delta v \sim 4.4~\kms$) are all characteristic of a SNR-MC interaction.

The northwestern cloud appears along the sharp boundary of the SNR shock front.  While this evidence may indicate physical contact between the SNR ejecta and molecular gas, there is little indication in $^{12}$CO $J=1-0$ emission to suggest an interaction scenario.  The cloud extends to the northwest of the remnant as part of a larger complex, its observed line width ($\Delta v \sim 2.7~\kms$) is only slightly larger than the surrounding gas, and no other signs of SNR-MC interaction appear in this region.

This potential interaction region may be supported by the presence of enhanced TeV gamma-ray emission, although the detection is just below the $5\sigma$ limit traditionally used to identify TeV sources (\autoref{tab:survey}).

In $^{12}$CO $J=2-1$, we detect the same molecular emission as seen in $^{12}$CO $J=1-0$ emission toward G16.7$+$0.1 at a systemic velocity of $+25~\kms$.  Toward the emission that \citet{rm00} identify as the ``southern cloud,'' some pixels appear velocity-broadened in $^{12}$CO $J=2-1$ to a width of around $\Delta v = 9.6~\kms$.  This value contrasts with the smaller velocity-width in $J=1-0$ of $1.5~\kms$ and suggests a recent or fast shock may be responsible for the large observed line widths.  While this width is broad compared to emission observed in $^{12}$CO $J=1-0$, it is also consistent with other shock indicators.  The southern cloud is the site of OH maser emission as well as a significant enhancement in the radio continuum.  We do not detect any broadened emission at the positions of the other clouds in this region.

\textit{Kes 75} was first observed to have a BML region in $^{12}$CO $J=1-0$ \citep{scyk09}.  A large MC is coincident with the northern half of the remnant and at a systemic velocity between $+45$ and $+58~\kms$.  The evidence for interaction is in the broadened blue wing of this feature ($+45$ to $+51~\kms$).  This asymmetric and broadened feature is spectroscopically similar to the SNR-MC interactions seen toward IC 443 and 3C 391.  Kes 75 is also the site of a strong TeV gamma-ray source \citep{djannati-atai+08}, although this emission is thought to be associated with a pulsar wind nebula in the remnant itself.

We confirm the BML detection in $^{12}$CO $J=2-1$ around $+53~\kms$.  Broadened molecular emission appears extended along the north and west of the remnant and is coincident with the brightest regions of 20 cm continuum emission.  This observation is also consistent with detection of broadened emission in $^{12}$CO $J=1-0$ where line widths were observed to be on the order of $3.7~\kms$.  In our analysis, we find broader line widths in $^{12}$CO $J=2-1$ up to $9.2~\kms$.

\textit{3C 391} is discussed in \autoref{sec:3c391}.

\textit{Kes 79} was observed to have a SNR-MC interaction in bright $^{12}$CO $J=1-0$ emission to the east and south of the radio shell of the remnant.  This morphological correspondence between the SNR and MC strongly implies an association between these objects \citep{swdf03}.  The CO line profile at these positions is centered around $+103~\kms$ with an integrated intensity of $75.9~\text{K km s}^{-1}$ and a peak brightness temperature of $\sim 8~\text{K}$.  Thus, the line has an approximate equivalent width of $\sim 9~\kms$, consistent with a shock-broadened cloud.  However, the observation of $^{12}$CO $J=1-0$ was performed with the NRAO $12$ m telescope and a beamwidth of $45''$.  The integrated intensity measurements reported in \citet{swdf03} are ambiguous and may include line emission from multiple, ambient MCs and not a single, shocked cloud.

We confirm detection of velocity-broadened $^{12}$CO emission in $J=2-1$ from $+99$ to $+109~\kms$ toward Kes 79.  Broad line detections extend across the MC from southwest toward the northeast of the remnant with velocity-widths up to $8.8~\kms$.  Agreement between our $^{12}$CO measurements and previous observations, as well as the OH measurements at similar velocities (\autoref{sec:anc}), argues for the presence of a SNR-MC interaction toward this remnant.

\textit{3C 396} is observed to have a possible SNR-MC interaction in $^{12}$CO $J=2-1$ and $J=1-0$ emission, where a bright cloud appears between $+67$ and $+72~\kms$ \citep{scyk11}.  At these velocities, the coincident MC extends around the remnant, with emission on the blue end of the line profile concentrated to the north and east and extending clockwise around the remnant toward redshifted emission along the western edge.  This detection represents strong morphological agreement between the structure of the MC and the SNR.  The spectral lines themselves exhibit signs of broadening, with an asymmetric profile and broad wings toward redder velocities in the eastern emission and blueshifted broad wings in the southwestern emission.  In some regions, these line profiles have widths $> 10~\kms$.  The presence of broadened molecular emission along the western edge of 3C 396 as well as the OH maser at the same velocity argues for the presence of a SNR-MC interaction.

We detect velocity-braodened $^{12}$CO $J=2-1$ emission centered around $+69~\kms$ to the north of the remnant and extending to the west where the broadened emission is centered around $+77~\kms$.  The detected broad line widths are generally between $6.6$ and $7.6~\kms$.  These broad-line detections are not bright or significantly enhanced compared to surrounding material.  While our analysis supports the presence of a SNR-MC interaction, detailed analysis of this region may reveal shocked material that is not significantly distinct from the surrounding gas.

Evidence for a SNR-MC interaction toward \textit{3C 397} comes from $^{12}$CO $J=1-0$ observations performed with the 13.7 m telescope at the Purple Mountain Observatory at Delingha \citep{jcws10}.  These data reveal a MC at $+32~\kms$ and to the north and west of the remnant.  At these positions, the SNR shock appears to be embedded in the MC and morphological correspondence between the $^{12}$CO and SNR radio continuum emission imply the objects are interacting.  The CO emission in these regions indicates that the MC emits a line profile with broad blue wings ($+28$ to $+31~\kms$), consistent with acceleration from a SNR-MC interaction.  The authors report a FWHM line width of $2.5~\kms$ toward the MC at this velocity.

We confirm detection of velocity-broadened emission along the western edge of the remnant and centered at a systemic velocity of $+31~\kms$.  The blue wing is not as bright in $^{12}$CO $J=2-1$ emission as in the lower-excitation line, although the overall line profile appears to be enhanced to $\Delta v = 6.2~\kms$.

Toward \textit{W49}, evidence for a SNR-MC interaction can be seen in $^{13}$CO $J=1-0$ emission \citep{simon01}.  Broad lines have been detected at a systemic velocities from $0$ to $+20~\kms$ in multiple components and with a narrower line profile centered at $+16~\kms$.  Averaged over the entirety of W49B, the line profile has a width of $\sim 10~\kms$, although the spatially resolved emission is likely narrower.

We detect broadened $^{12}$CO $J=2-1$ emission toward W49B in the same velocity range and centered at a systemic velocity of $+14~\kms$.  Regions of broadened emission are seen to the west of W49B, and the majority of our detections come from a cloud located along the southwestern boundary of the remnant.  In this region, we report BML detections with widths of $\Delta v = 10~\kms$.  

While this finding agrees with previous detections of broadened molecular emission, some interpretations argue against the presence of a SNR-MC interaction to the southwest of W49B.  \citet{lacey+01} found that low-frequency radio emission in this region is almost completely attenuated, and the intervening molecular material responsible for absorption may only be seen toward the remnant in projection.  However, at higher frequencies W49B becomes significantly brighter.  Indeed, at 90, 20, and 6 cm, the southwestern portions of the shell show significant local enhancement \citep{moffett+94}.  Added to this fact is the observation that spectral indices appear to flatten to the southwest of W49B.  It may be that this region is not enhanced in nonthermal emission.  Overall, the evidence for the presence of a SNR-MC interaction is ambiguous, though our observations support the presence of an interaction scenario.  Detailed characterization of the physical properties in the gas itself is needed to further verify the presence of a SNR-MC interaction.

\textit{Cas A} is discussed in \autoref{sec:casa}.

Evidence for a potential SNR-MC interaction toward \textit{HB 3} was reported by \citet{routledge91}, in which the authors perform \HI\ line mapping and $^{12}$CO $J=1-0$ observations.  Toward this remnant, a bright $^{12}$CO ``bar'' is coincident with a region of enhanced radio emission near $l = +134^{\circ}, b = +1^{\circ}$ and extending east in Galactic coordinates.  This morphological correspondence suggests the SNR is interacting with the cloud and a region of enhanced particle acceleration is seen in the SNR ejecta.  Evidence from \citet{shi+08} supports this hypothesis where radio continuum from $1.4$ to $4.8~\text{GHz}$ is brighter toward the $^{12}$CO cloud.  The spectral indices at these frequencies appear to steepen to $\alpha = -0.61$.  Emission from the MC itself extends to the south of the remnant, roughly in the direction of the OH maser detection, and the line profile peaks at a velocity of $-43~\kms$.  

However, the W3 \HII\ region is embedded in a MC at $-40.5~\kms$, and it is certain that the $^{12}$CO ``bar'' at $-43~\kms$ is associated with this cloud.  \citet{bp11} detect bright $^{12}$CO $J=2-1$ and $J=3-2$ emission centered on W3(OH) in this same region.  The correlation between the \HII\ region, high line-ratio $^{12}$CO emission, and 1.1 mm dust continuum from the Bolocam Galactic Plane Survey \citep{rosolowsky+10} argues that these BML regions are not associated with HB 3.

We detect $^{12}$CO $J=2-1$ emission in this region and centered at a systemic velocity of $-40~\kms$.  Our spectral analysis confirms that this emission is broadened, up to $\Delta v = 12.7~\kms$, although generally to a velocity width of $\Delta v = 7.6$ to $8~\kms$ in $^{12}$CO $J=2-1$.

\textit{IC 443} is a well-observed SNR with several indications of a SNR-MC interaction.  This interaction was first identified in broadened CO emission by \citet{cornett77} and has been confirmed by detection of molecular species toward the associated MCs \citep{white86, xwm11}, OH $1720~\text{MHz}$ emission \citep{claus97, hew06}, and H$_{2}$ emission \citep{den79a, den79b, burt88}.  The physical conditions toward IC 443 are well-suited for every shock interaction tracer traditionally used to identify SNR-MC interactions.  As a result, this region is one of the most luminous molecular hydrogen sources in the galaxy.  The interacting MC generally exhibits linewidths of $> 30~\kms$, up to $90~\kms$ toward the most turbulent molecular gas \citep{white86}.  This result has been confirmed in several molecules, probing molecular gas of densities up to $3 \times 10^{6}$ cm$^{-3}$ \citep{vjp93}.

The SNR itself has a composite morphology and appears evolved, with an inferred age of $\sim 2 \times 10^{4}$ \citep{leahy+04,lee+08}.  The overall radio morphology of the remnant was described by \citet{bs86} in $327~\text{MHz}$ continuum emission.  Two main subshells (shells A and B) make up the majority of the remnant in radio continuum emission.  Shell A appears brighter and is coincident with the molecular shock indicators along its southern boundary and across the center of the remnant as a whole.  The western edge of this shell appears to be associated with both the OH maser emission and TeV gamma-rays \citep{albert+07b}.

Given its approximate age and the assumed distance of $\sim 1.5~\text{kpc}$, IC 443 has a large apparent size on the order of $1^{\circ}$.  We targeted a section of the remnant toward its center and along the southwestern edge of shell A.  The shocked $^{12}$CO emission peaks around $-17~\kms$ in this region and extends blueward in a broad, asymmetric line profile.  Our observations centered on a bright cloud referred to as shock clump B in \citet{den79a}.  We measured velocity-widths in $^{12}$CO $J=2-1$ toward this region ranging from $\Delta v = 7.6$ to $28~\kms$.  

\subsection{Non-Detection Toward G54.1$+$0.3}\label{sec:non}

G54.1$+$0.3 was targeted with $^{12}$CO $J=2-1$ observations in \citet{lee+12}.  A large MC at $+23~\kms$ extends along the north of the remnant, with a bright central core to the northeast of the remnant.  The morphological correspondence between the outer blastwave of the SNR and the MC is not strong, with at least $30''$ separation between the 20 cm contours of the remnant and the MC.  \citet{lee+12} report a FWHM of $7~\kms$ at $+23~\kms$ across the entire cloud.

\begin{figure*}
	\begin{center}
		\begin{minipage}[t][][b]{0.42\textwidth}
			\centering
			\includegraphics[width=\textwidth,angle=270,origin=c]{G541+03_region01.eps}
		\end{minipage}
		\hfill
		\begin{minipage}[t][][b]{0.42\textwidth}
			\centering
			\includegraphics[width=\textwidth]{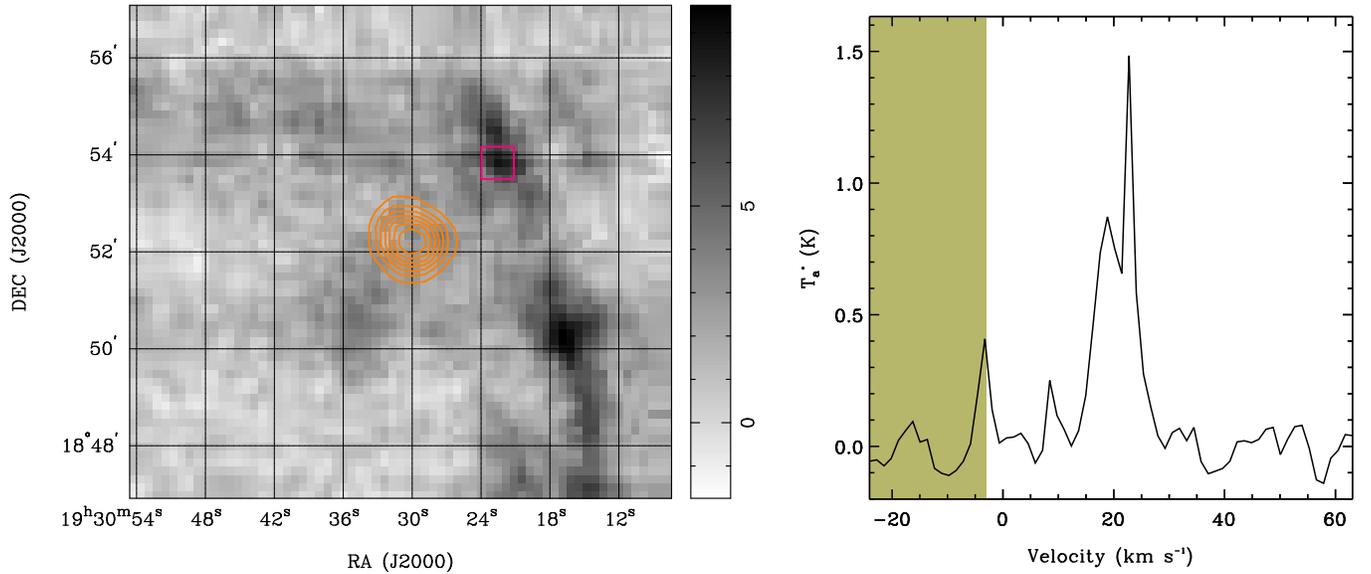}
		\end{minipage}
	\end{center}
	\vspace{-0.5in}
		\caption{\scriptsize (Left) Integrated $^{12}$CO $J=2-1$ emission toward G54.1$+$0.3 from $+16$ to $+23~\kms$.  Orange contours are NVSS $1.4~\text{GHz}$ continuum emission of the remnant.  Pixels with broadened molecular emission would be shown with white squares, but none passed our criteria.  We denote a region of the $+23~\kms$ cloud discussed in \citet{lee+12} with a red square.  (Right) The average spectrum across the pixels denoted by the red square.  There appear to be two line components centered at $+20$ and $+23~\kms$.  Neither component exhibits broadening at the $> 6~\kms$ level we used in our analysis.}\label{fig:G541+03}
\end{figure*}

Our observations indicate that the broad feature detected around $+19~\kms$ with $\Delta v \sim 11~\kms$ \citep[see Figure 13, therein]{lee+12} is at least two clouds blended over the $30''$ grid spacing applied in this study.  While the broad and narrow components of emission at these velocities can be separated from $+15$ to $+30~\kms$, the broad component is further separated into two features centered at $+20$ and $+23~\kms$ (\autoref{fig:G541+03}).  These components are resolved in our spectra.  However, these features may have been smoothed into a single, broad-line component in \citet{lee+12} given the larger beam ($48''$) from the 6-m Seoul Radio Astronomy Observatory Telescope.  In contrast to the Gaussian component analysis over a large region, our pixel-by-pixel analysis of the molecular emission reveals no indication of $> 6~\kms$ velocity-broadening.

\subsection{Newly Identified SNR-MC Systems}\label{sec:new}

We detected broadened molecular features toward nine SNRs with no previous indication of SNR-MC interactions: G08.3$-$0.0, G09.9$-$0.8, G11.2$-$0.3, G12.2$+$0.3, G18.6$-$0.2, G23.6$+$0.3, 4C$-$04.71, G29.6$+$0.1, G32.4$+$0.1.  In this section, we discuss the characteristics of the interaction as identified by our BML region algorithm along with previous work on the associated SNRs.  Our results from the detection of velocity-broadened regions toward each remnant are summarized in \autoref{tab:new}.  For comparison with each remnant, we use 90 cm radio continuum contours from \citet{bggk06} (hereafter, BGGKL06) where $l < +22^{\circ}$ and NVSS $1.4~\text{GHz}$ contours \citep{condon+98} for the remaining objects.

\subsubsection{G08.3$-$0.0}\label{sec:G083-00}

G08.3$-$0.0 was initially identified in 20 cm emission by \citet{helfand06} and subsequently confirmed by BGGKL06 at 90 cm as a small ($5' \times 4'$) SNR with shell-like morphology.  The remnant has flux density $1.0~\text{Jy}$ at 20 cm, but its morphology suggests an enhancement in surface brightness toward the east.  A subsequent search for associated OH maser emission at 1612, 1665, 1667, and $1720~\text{MHz}$ with the Green Bank Telescope (GBT) yielded a single-dish detection with no velocity information or localization \citep{hy09}.  Of particular note is a TeV gamma-ray detection reported toward this remnant \citep{higashi+08}.  The gamma-ray emission was observed with a $7'$ offset from the center of G08.3$-$0.0 and detected with $\sim 14'$ resolution.  Given the position of the remnant relative to the gamma-ray detection, there is reason to believe that the gamma-ray source originates from the northern half of this SNR.  

\begin{figure*}
	\begin{center}
		\begin{minipage}[t][][b]{0.42\textwidth}
			\centering
			\includegraphics[width=\textwidth,angle=270,origin=c]{G083-00_region04.eps}
		\end{minipage}
		\hfill
		\begin{minipage}[t][][b]{0.42\textwidth}
			\centering
			\includegraphics[width=\textwidth]{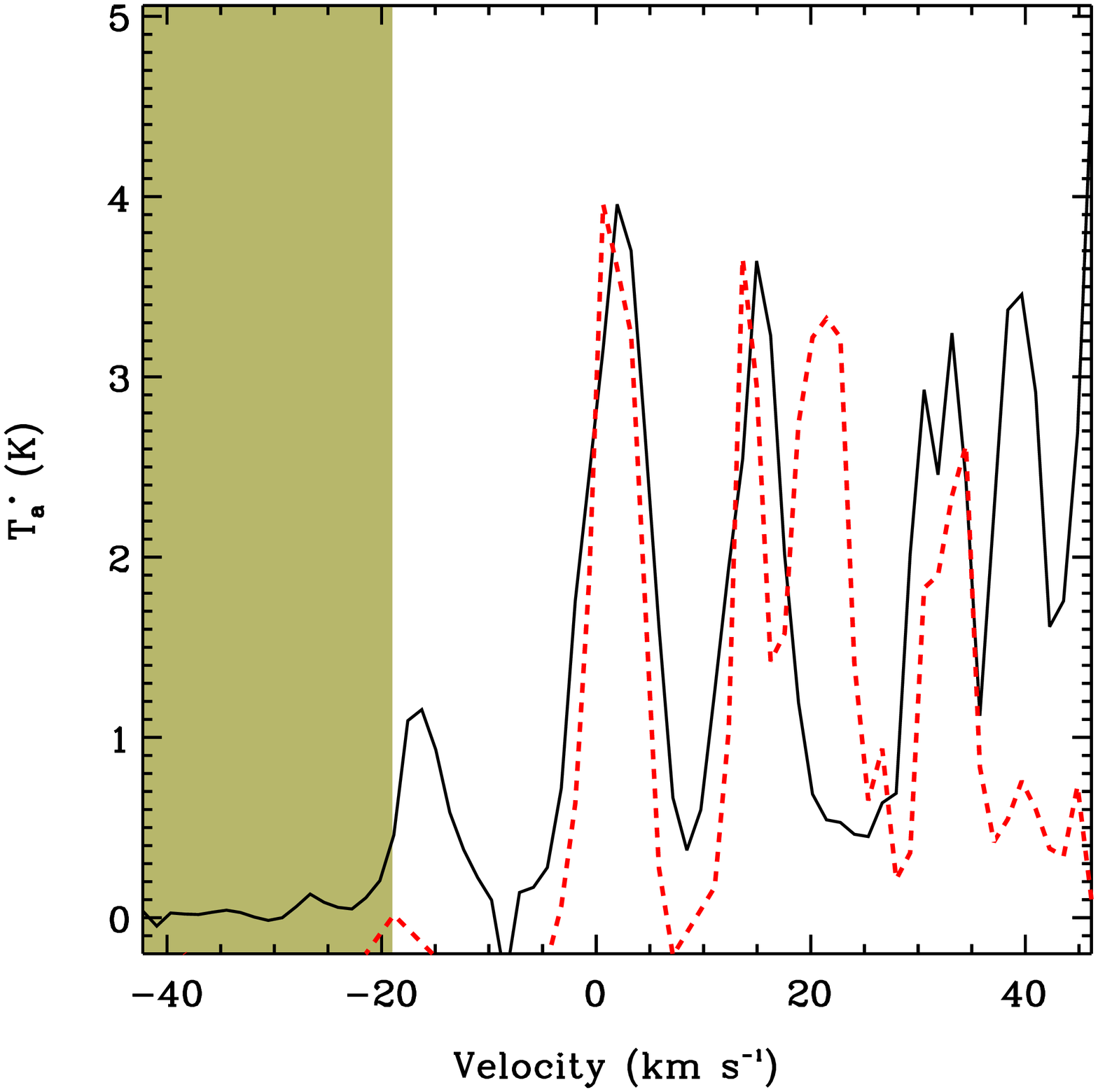}
		\end{minipage}
	\end{center}
	\vspace{-0.5in}
		\caption{\scriptsize (Left) Integrated $^{12}$CO $J=2-1$ emission toward G08.3$-$0.0 from $-2$ to $+6~\kms$.  Orange contours are 90 cm radio continuum emission of the remnant from BGGKL06.  Pixels with broadened molecular emission are shown as white squares.  We denote a comparison region of unbroadened, bright molecular emission with a larger red square.  (Right) The average spectrum from these pixels at the velocity of the BML identification (solid black) along with a spectrum of the unbroadened, bright comparison region averaged over the pixels denoted by the red square (dashed red).  Shaded parts of the spectrum correspond to velocities ruled out by the inferred distance to the SNR.}\label{fig:G083-00}
\end{figure*}

In projection, the radio contours of G08.3$-$0.0 appear to flatten toward the northwest (\autoref{fig:G083-00}) coincident with a MC between $-0.7$ and $4.6~\kms$.  The spectrum of the molecular emission in this region appears velocity-broadened with $^{12}$CO Gaussian widths fit between $6.5$ and $9.0~\kms$.  The coincidence of this broadened $^{12}$CO along with the morphology of the remnant, the possible OH maser toward this position, and the gamma-ray source associated with this region support the presence of a SNR-MC interaction at this location at a systematic velocity near $+2.6~\kms$.  We note that the velocity centroid of the BML region and the kinematic distance inferred from this velocity, $d = 16.4~\text{kpc}$, is also in good agreement with the value given by the $\Sigma-D$ relation ($16.3~\text{kpc}$).

\subsubsection{G09.9$-$0.8}

G09.9$-$0.8 has been observed to have typical radio shell-like morphology (BGGKL06).  Its spectral index between $90$ and $20$ cm is Crab-like ($\alpha = -0.4$) implying nonthermal emission from the SNR.  There may also be confusion with thermal emission from other sources, which suggests that the SNR has a steeper spectral profile in agreement with enhanced particle acceleration.  H$\alpha$ emission toward the remnant is correlated with nonthermal emission at $1.4~\text{GHz}$ along the northern edge of the remnant \citep{sp11}.  This trend implies a connection between regions of strong SNR-ISM interaction and nonthermal emission from the SNR along that edge of the shell.

\begin{figure*}
	\begin{center}
		\begin{minipage}[t][][b]{0.38\textwidth}
			\centering
			\includegraphics[width=\textwidth,angle=270,origin=c]{G099-08_region01.eps}
		\end{minipage}
		\hfill
		\begin{minipage}[t][][b]{0.38\textwidth}
			\centering
			\includegraphics[width=\textwidth]{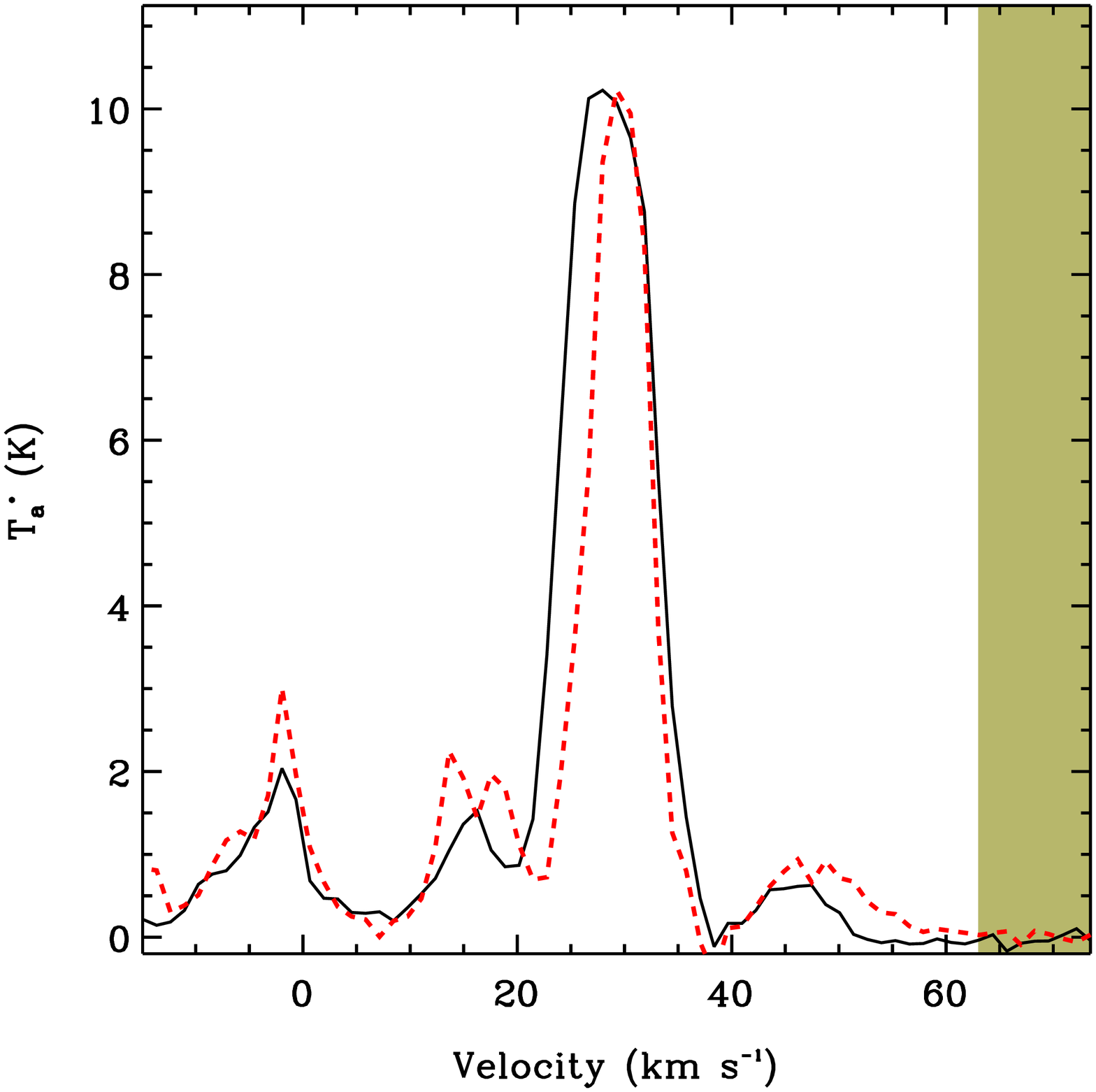}
		\end{minipage}
	\end{center}
	\vspace{-0.5in}
		\caption{\scriptsize (Left) Integrated $^{12}$CO $J=2-1$ emission toward G09.9$-$0.8 from $+26$ to $+34~\kms$.  Orange contours are 90 cm radio continuum emission of the remnant from BGGKL06.  Pixels with broadened molecular emission are shown as white squares.  We denote a comparison region of unbroadened, bright molecular emission with a larger red square.  (Right) The average spectrum from these pixels at the velocity of the BML identification (solid black) along with a spectrum of the unbroadened, bright comparison region averaged over the pixels denoted by the red square (dashed red).  Shaded parts of the spectrum correspond to velocities ruled out by the inferred distance to the SNR.}\label{fig:G099-08}
\end{figure*}

We observe an association between a bright region of 90 cm radio emission and a MC toward the southwest of this remnant (\autoref{fig:G099-08}).  The molecular emission appears at a systemic velocity of $+27$ to $+33~\kms$ and extends in a small arc toward the center of the remnant.  The region of velocity-broadened emission is concentrated toward a peak in the 90 cm contours, and the molecular emission to the east of this peak exhibits the most velocity-broadening.  The $^{12}$CO Gaussian widths of the velocity-broadened emission in this region are fit between $6.6$ and $9.1~\kms$.

\subsubsection{G11.2$-$0.3}

G11.2$-$0.3 was identified as a SNR by \citet{sg70} at 408 and $5000~\text{MHz}$.  These observations revealed a small remnant ($4'$) with a steep spectral index of $\alpha = -0.52$.  Follow-up with ASCA indicated there is a 65 ms pulsar toward the center of the remnant \citep{torii+97}.  Subsequent radio detection of the remnant in BGGKL06 revealed a probable shell-like SNR with a flux density at 20 cm of approximately $0.6~\text{Jy}$.  Emission from the remnant at 90 cm appears locally enhanced around the shell, and especially to the east and northwest.  Detection of OH maser emission has been reported from GBT observations \citep{green97}, however, the lack of precise position and velocity information makes it difficult to associate the OH maser with the SNR.  There appears to be a strong TeV gamma-ray source extended to the northwest of the remnant and HESS detects emission toward G11.2$-$0.3 at $5.5\sigma$ \citep{bochow11}.

\begin{figure*}
	\begin{center}
		\begin{minipage}[t][][b]{0.42\textwidth}
			\centering
			\includegraphics[width=\textwidth,angle=270,origin=c]{G112-03_region01.eps}
		\end{minipage}
		\hfill
		\begin{minipage}[t][][b]{0.42\textwidth}
			\centering
			\includegraphics[width=\textwidth]{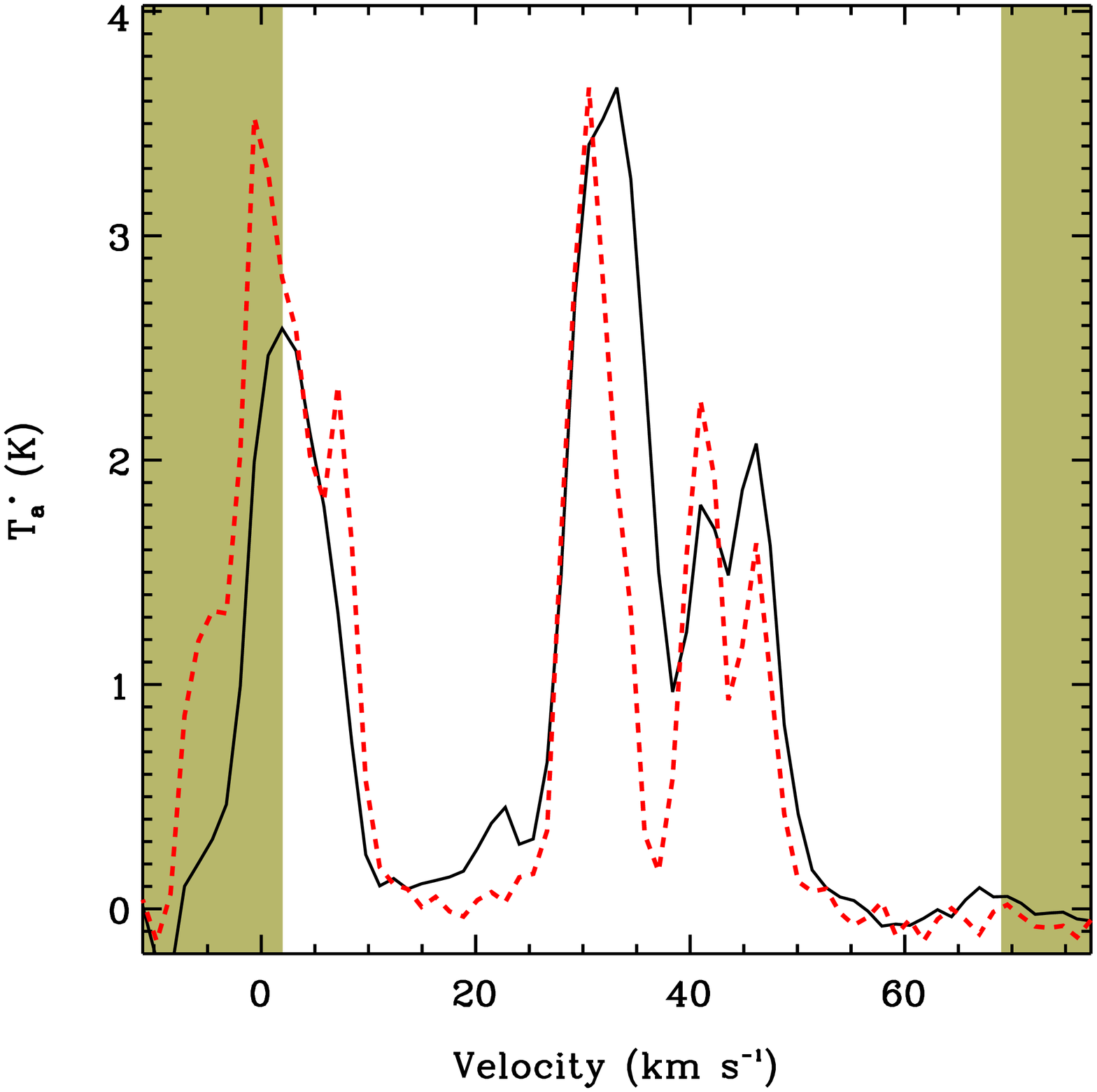}
		\end{minipage}
	\end{center}
	\vspace{-0.5in}
		\caption{\scriptsize (Left) Integrated $^{12}$CO $J=2-1$ emission toward G11.2$-$0.3 from $+30$ to $+38~\kms$.  Orange contours are 90 cm radio continuum emission of the remnant from BGGKL06.  Pixels with broadened molecular emission are shown as white squares.  We denote a comparison region of unbroadened, bright molecular emission with a larger red square.  (Right) The average spectrum from these pixels at the velocity of the BML identification (solid black) along with a spectrum of the unbroadened, bright comparison region averaged over the pixels denoted by the red square (dashed red).  Shaded parts of the spectrum correspond to velocities ruled out by the inferred distance to the SNR.}\label{fig:G112-03}
\end{figure*}

We detect velocity-broadened molecular emission toward this remnant from $+29$ to $+36~\kms$.  At these velocities, emission extends from the southeast to the north of the remnant, roughly coincident with the bright shell of G11.2$-$0.3 as seen in 90 cm contours (\autoref{fig:G112-03}).  Regions of velocity-broadened emission lie mostly toward the northern molecular emission, with some detections interior to the main radio shell.  The Gaussian line width of velocity-broadened emission in this region is between $6.6$ and $9.1~\kms$.

\subsubsection{G12.2$+$0.3}

G12.2$+$0.3 was discovered in BGGKL06 as a small, partial shell-like remnant.  The radio spectral index from 90 to 20 cm is steep with $\alpha \sim -0.7$.  The weak flux density at 20 cm combined with the size suggest that the remnant has $d > 10~\text{kpc}$, and the $\Sigma-D$ relationship implies a distance of $15.6$ kpc.  $1720~\text{MHz}$ OH maser emission has been detected toward this remnant in single-dish observations with no reported velocity information or localization \citep{hy09}.

\begin{figure*}
	\begin{center}
		\begin{minipage}[t][][b]{0.42\textwidth}
			\centering
			\includegraphics[width=\textwidth,angle=270,origin=c]{G122+03_region01.eps}
		\end{minipage}
		\hfill
		\begin{minipage}[t][][b]{0.42\textwidth}
			\centering
			\includegraphics[width=\textwidth]{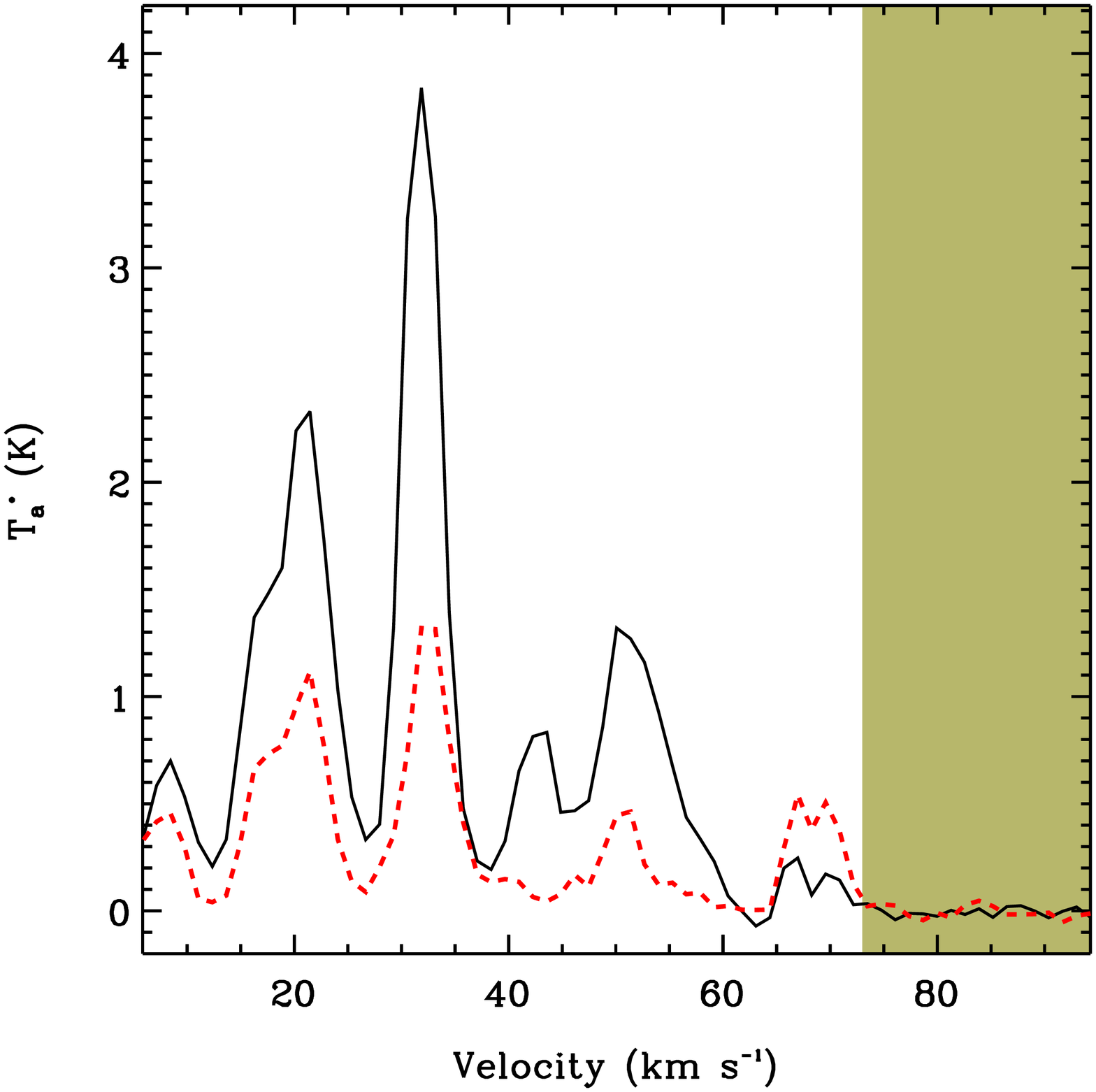}
		\end{minipage}
	\end{center}
	\vspace{-0.5in}
		\caption{\scriptsize (Left) Integrated $^{12}$CO $J=2-1$ emission toward G12.2$+$0.3 from $+48$ to $+55~\kms$.  Orange contours are 90 cm radio continuum emission of the remnant from BGGKL06.  Pixels with broadened molecular emission are shown as white squares.  We denote a comparison region of unbroadened, bright molecular emission with a larger red square.  (Right) The average spectrum from these pixels at the velocity of the BML identification (solid black) along with a spectrum of the unbroadened, bright comparison region averaged over the pixels denoted by the red square (dashed red).  Shaded parts of the spectrum correspond to velocities ruled out by the inferred distance to the SNR.}\label{fig:G122+03}
\end{figure*}

We detect velocity-broadened $^{12}$CO emission coincident with this remnant at a systemic velocity of $+50~\kms$ (\autoref{fig:G122+03}).  The broadened molecular emission lies toward the eastern shock front of the remannt and to the north of the brightest region in 90 cm continuum emission.  This emission is weakly detected in two pixels of our map, and the velocity-width between the two BML detections extends from $6.6$ to $7.2~\kms$.  However, the morphology of the coincident MC and SNR are suggestive of an interaction at this velocity.  As seen in projection, the shape of the SNR appears flattened to the east and along the boundary of the SNR.  This shape contrasts with the northwest edge of the SNR where the shell structure appears much fainter and more extended.  We infer that the eastern shock front is deccelerated relative to the rest of the remnant, which would require a significant density enhancement in the ISM.

\subsubsection{G18.6$-$0.2}

\begin{figure*}
	\begin{center}
		\begin{minipage}[t][][b]{0.38\textwidth}
			\centering
			\includegraphics[width=\textwidth,angle=270,origin=c]{G186-02_region01.eps}
		\end{minipage}
		\hfill
		\begin{minipage}[t][][b]{0.38\textwidth}
			\centering
			\includegraphics[width=\textwidth]{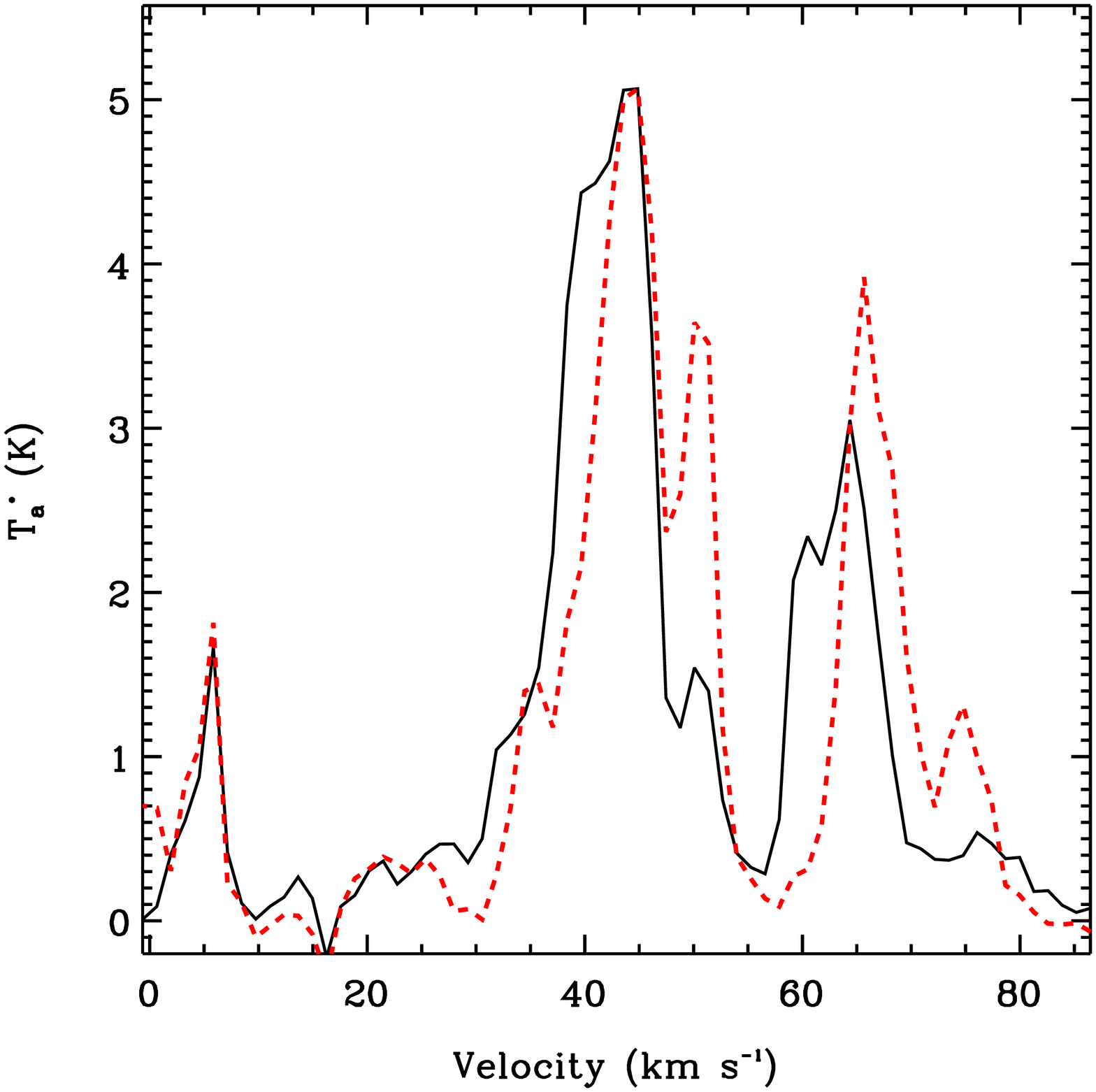}
		\end{minipage}
	\end{center}
	\vspace{-0.6in}
		\caption{\scriptsize (Left) Integrated $^{12}$CO $J=2-1$ emission toward G18.6$-$0.2 from $+40$ to $+48~\kms$.  Orange contours are 90 cm radio continuum emission of the remnant from BGGKL06.  Pixels with broadened molecular emission are shown as white squares.  We denote a comparison region of unbroadened, bright molecular emission with a larger red square.  (Right) The average spectrum from these pixels at the velocity of the BML identification (solid black) along with a spectrum of the unbroadened, bright comparison region averaged over the pixels denoted by the red square (dashed red).}\label{fig:G186-02}
\end{figure*}

G18.6$-$0.2 was first confirmed as a SNR by BGGKL06 in 90 cm continuum emission, although \citet{helfand06} listed the source G18.6375-0.2917 as a ``high-probability supernova remnant candidate'' before this confirmation. Subsequent detection at 20 cm identified a partial shell with a spectral index of $\alpha = -0.3$.  The remnant is small ($\sim 6'$) and faint in 90 cm emission ($1.9~\text{Jy}$), in agreement with a far distance as determined from the $\Sigma-D$ relation ($13.2$ kpc).  As shown in \autoref{fig:G186-02}, the region of brightest radio continuum emission toward this remnant extends in an arc to the east with fainter lobes of emission to the southwest and north.

We find evidence for a SNR-MC interaction to the north of the remnant at a systemic velocity of $+44~\kms$.  The MC in this region covers most of the northern and western half of the SNR, although the disturbed region appears localized to the north.  The velocity-widths of emission lines detected toward this cloud are broad, with most lines exceeding $\Delta v = 8.0~\kms$.

\subsubsection{G23.6$+$0.3}

It has been suggested based on its high infrared to radio luminosity ratio that G23.6$+$0.3 is an \HII\ region and not a SNR \citep{pnpm11}.  However, this object has a Crab-like spectral index ($\alpha = -0.34$), which implies some nonthermal component to the emission from this source \citep{sg70}.  While the radio emission is elongated, it correlates well with the spatial extent of the infrared emission and suggests that the two are associated.  This observation is consistent with the possibility that G23.6$+$0.3 is an \HII\ region and SNR projected onto each other.

\begin{figure*}
	\begin{center}
		\begin{minipage}[t][][b]{0.38\textwidth}
			\centering
			\includegraphics[width=\textwidth,angle=270,origin=c]{G236+03_region03.eps}
		\end{minipage}
		\hfill
		\begin{minipage}[t][][b]{0.38\textwidth}
			\centering
			\includegraphics[width=\textwidth]{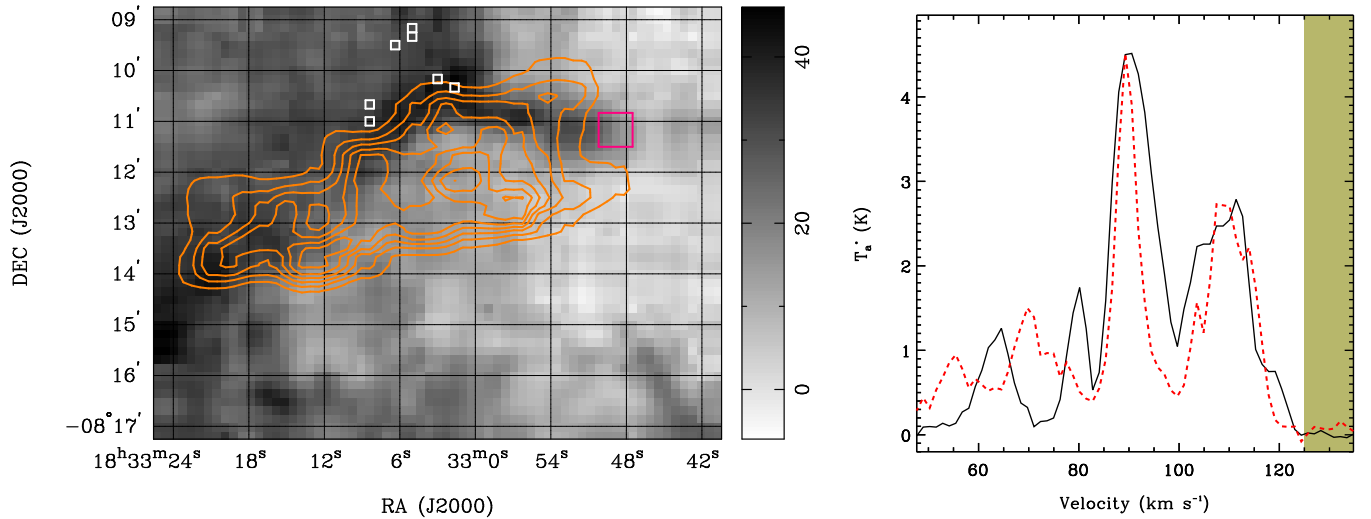}
		\end{minipage}
	\end{center}
	\vspace{-0.5in}
		\caption{\scriptsize (Left) Integrated $^{12}$CO $J=2-1$ emission toward G23.6$+$0.3 from $-43$ to $-35~\kms$.  Orange contours are NVSS $1.4~\text{GHz}$ continuum emission of the remnant.  Pixels with broadened molecular emission are shown as white squares.  We denote a comparison region of unbroadened, bright molecular emission with a larger red square.  (Right) The average spectrum from these pixels at the velocity of the BML identification (solid black) along with a spectrum of the unbroadened, bright comparison region averaged over the pixels denoted by the red square (dashed red).  Shaded parts of the spectrum correspond to velocities ruled out by the inferred distance to the SNR.}\label{fig:G236+03}
\end{figure*}

We detect velocity-broadened lines toward the remnant at a systemic velocity of $+91~\kms$.  The regions of broadened molecular emission lie along the brightest part of the remnant's northern shell (\autoref{fig:G236+03}).  Emission in this region appears moderately broadened, and the emission lines have velocity-widths around $\Delta v \sim 6.5~\kms$ up to $7.5~\kms$.  As seen in projection, the morphology of the SNR relative to the MC is highly suggestive.  There appears to be a boundary along the northern edge of G23.6$+$0.3 where an ongoing interaction has broadened the coincident MC.

\subsubsection{4C$-$04.71}

4C$-$04.71 is often referred to as Kes 73 and was identified as a potential SNR in part of a larger radio source centered around Galactic coordinates $l = +27.3, b=0.0$ by \citet{milne69}.  The remnant is compact with a central pulsar identified at X-ray wavelengths and an age of $\sim 2000~\text{yr}$ \citep{vg97}.  \citet{green97} were able to localize OH maser emission toward this source $12'$ west of the SNR around $+33~\kms$.  Subsequent observations revealed associated $^{13}$CO $J = 1 - 0$ and \HI\ emission toward this source and two nearby \HII\ regions imply a connection with gas around $+110~\kms$ \citep{tl08}.  HESS detected TeV gamma-ray emission toward this remnant at $6\sigma$ significance \citep{aharonian+08b}.  There appears to be a TeV source extending to the south of the remnant, although several other counterparts have been discussed for TeV detections in this region.

\begin{figure*}
	\begin{center}
		\begin{minipage}[t][][b]{0.42\textwidth}
			\centering
			\includegraphics[width=\textwidth,angle=270,origin=c]{4C-0471_region01.eps}
		\end{minipage}
		\hfill
		\begin{minipage}[t][][b]{0.42\textwidth}
			\centering
			\includegraphics[width=\textwidth]{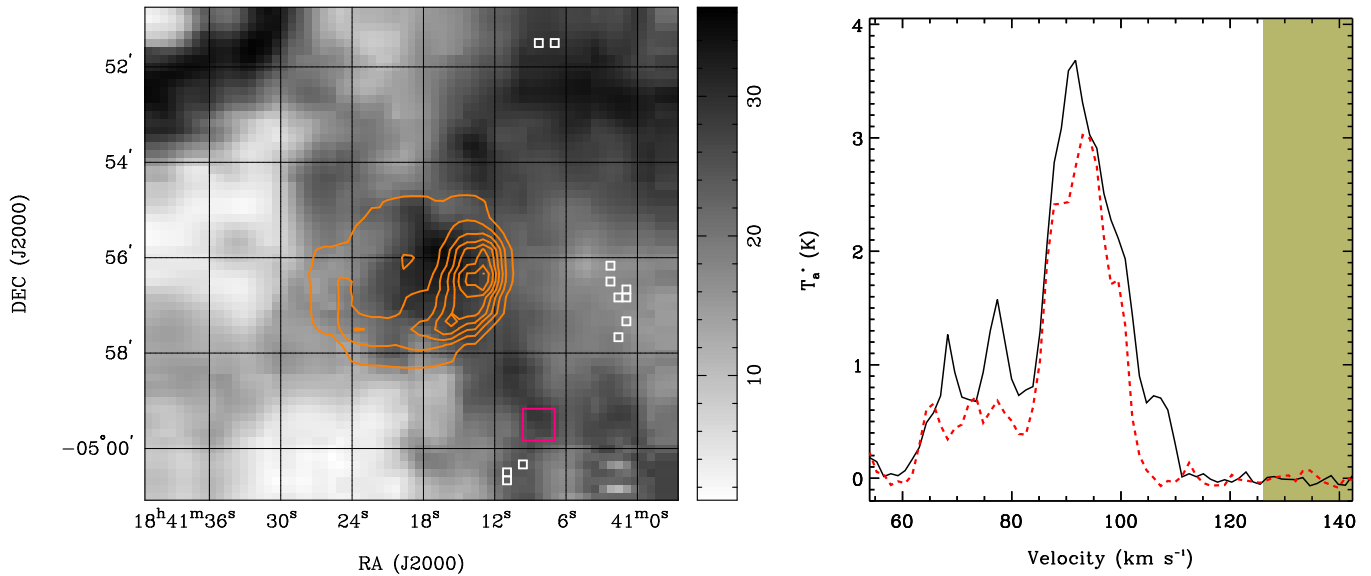}
		\end{minipage}
	\end{center}
	\vspace{-0.5in}
		\caption{\scriptsize (Left) Integrated $^{12}$CO $J=2-1$ emission toward 4C$-$04.71 from $+94$ to $+102~\kms$.  Orange contours are NVSS $1.4~\text{GHz}$ continuum emission of the remnant.  Pixels with broadened molecular emission are shown as white squares.  We denote a comparison region of unbroadened, bright molecular emission with a larger red square.  (Right) The average spectrum from these pixels at the velocity of the BML identification (solid black) along with a spectrum of the unbroadened, bright comparison region averaged over the pixels denoted by the red square (dashed red).  Shaded parts of the spectrum correspond to velocities ruled out by the inferred distance to the SNR.}\label{fig:4c-0471-2}
\end{figure*}

We observe broadening from the same molecular emission toward 4C$-$04.71 as seen in \citet{tl08}, although in $^{12}$CO $J=2-1$ the line profile of the broadened emission is centered around $+100~\kms$ with average line widths of $\Delta v = 9.0~\kms$ (\autoref{fig:4c-0471-2}).  However, the broadened regions lie at large separations from 4C$-$04.71 ($\sim 3'$).  It is possible that these BML regions toward 4C$-$04.71 are associated with other nearby radio sources \citep[see, e.g.,]{sh92} often assumed to be \HII\ regions.  These sources would need to extend to within $3'$ in projection toward 4C$-$04.71 to its west and southwest, which is not seen in the radio continuum.  

A second possibility, which has been raised in the context of Cas A \citep{kilpatrick+14}, is that faint, fast-moving ejecta from 4C$-$04.71 could interact with the molecular gas without being visible in radio emission.  If ejected at high-velocity during early phases of SN evolution, this material could carry enough momentum to turbulently accelerate the MC to the velocity width we observe.  This hypothesis is supported by the observation that 4C$-$04.71 is still relatively young and the turbulent clouds are nearby as seen in projection.  Presumably some fraction of the ejecta has not yet been significantly deccelerated by the surrounding ISM and would not need to travel far to shock the observed MCs.

\subsubsection{G29.6$+$0.1}

G29.6$+$0.1 was discovered in VLA observations at 6 and 3.7 cm by \citet{gaensler+99}.  The radio source is a faint, shell-like remnant, and several local enhancements were detected in the continuum.  The spectral index from 6 to 3.7 cm is roughly $\alpha \sim -0.5$ across the remnant, although the brightest region of 6 cm continuum toward the eastern shell is steeper, with $\alpha = -0.77$.  The northwestern edge of the shell also appears relatively bright, although there is no evidence of steeper nonthermal emission.

Subsequent observations toward G29.6$+$0.1 with ASCA revealed a compact X-ray source thought to be a pulsar \citep{vgtg00}.  The X-ray source is relatively faint, and no timing or age estimate have been performed, although the coincidence of this source strongly implies that G29.6$+$0.1 is a distant and possibly young SNR from a core-collapse SN.

\begin{figure*}
	\begin{center}
		\begin{minipage}[t][][b]{0.42\textwidth}
			\centering
			\includegraphics[width=\textwidth,angle=270,origin=c]{G296+01_region01.eps}
		\end{minipage}
		\hfill
		\begin{minipage}[t][][b]{0.42\textwidth}
			\centering
			\includegraphics[width=\textwidth]{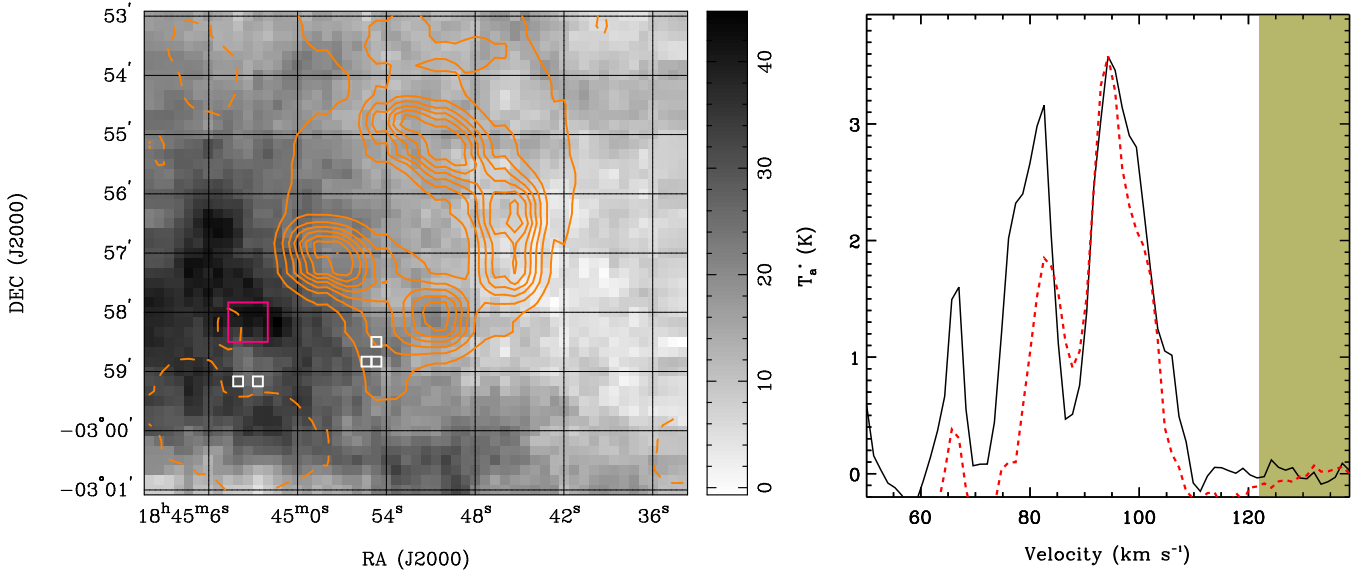}
		\end{minipage}
	\end{center}
	\vspace{-0.5in}
		\caption{\scriptsize (Left) Integrated $^{12}$CO $J=2-1$ emission toward G29.6$+$0.1 from $+34$ to $+42~\kms$.  Orange contours are NVSS $1.4~\text{GHz}$ continuum emission of the remnant.  Pixels with broadened molecular emission are shown as white squares.  We denote a comparison region of unbroadened, bright molecular emission with a larger red square.  (Right) The average spectrum from these pixels at the velocity of the BML identification (solid black) along with a spectrum of the unbroadened, bright comparison region averaged over the pixels denoted by the red square (dashed red).  Shaded parts of the spectrum correspond to velocities ruled out by the inferred distance to the SNR.}\label{fig:G296+01}
\end{figure*}

We detected broadened $^{12}$CO emission toward G29.6$+$0.1 and roughly coincident with a region of bright 20 cm continuum emission, at a velocity of $+94~\kms$ (\autoref{fig:G296+01}).  This position is consistent with enhancements in the 6 cm continuum and spectral index, as well as a region of bright X-ray emission as mapped by \citet{vgtg00}.  The majority of the molecular emission at these velocities coincides with the remnant, implying a strong morphological association between the SNR and MCs.  A weak red wing is also visible in the $^{12}$CO spectrum demonstrating the characteristic velocity profile of shocked-broadened molecular emission.  The two pixels where we find evidence for velocity-broadened molecular emission are highly disturbed, with line widths $\Delta v \sim 10.0~\kms$.

\subsubsection{G32.4$+$0.1}

\begin{figure*}
	\begin{center}
		\begin{minipage}[t][][b]{0.42\textwidth}
			\centering
			\includegraphics[width=\textwidth,angle=270,origin=c]{G324+01_region02.eps}
		\end{minipage}
		\hfill
		\begin{minipage}[t][][b]{0.42\textwidth}
			\centering
			\includegraphics[width=\textwidth]{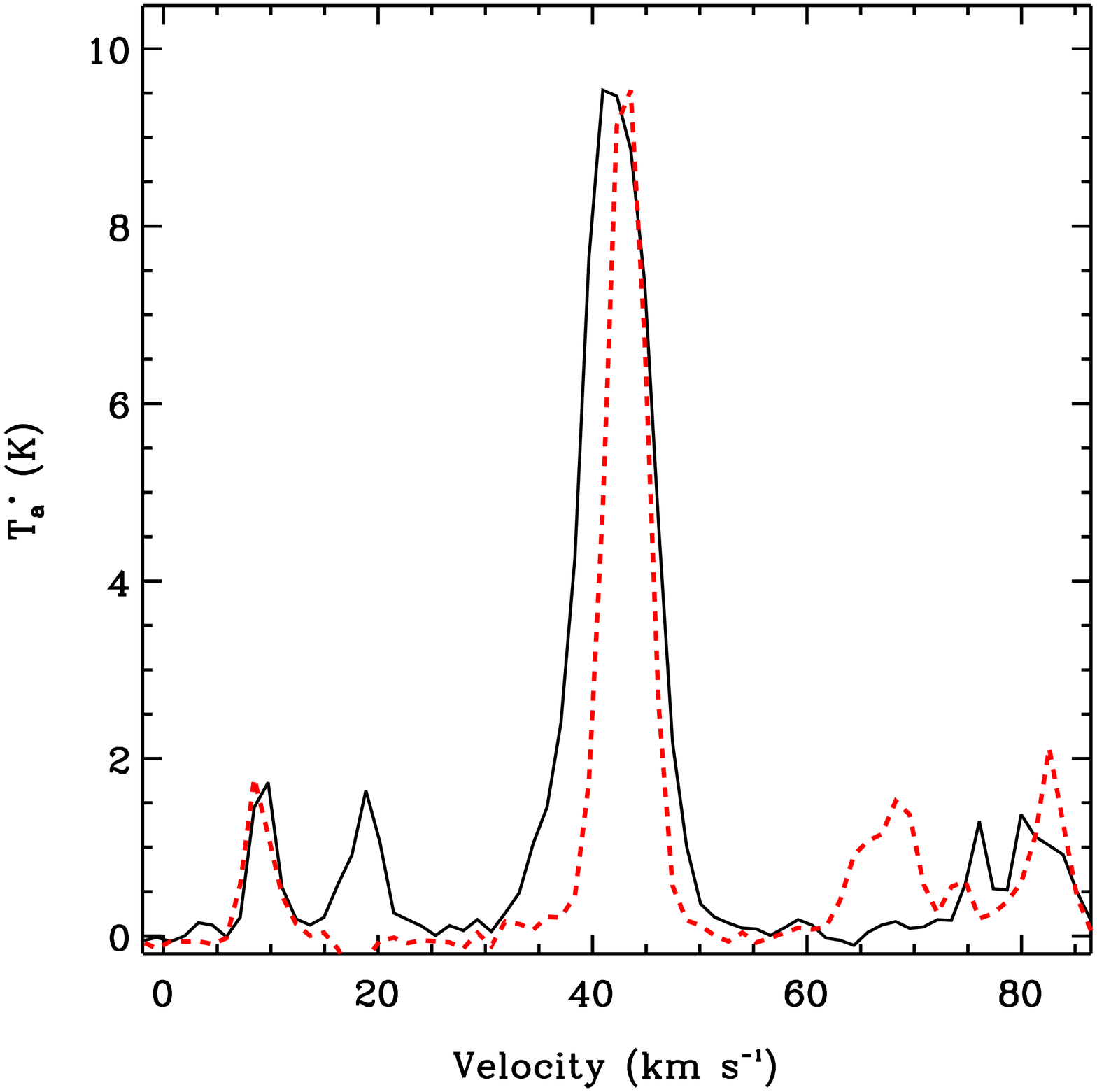}
		\end{minipage}
	\end{center}
	\vspace{-0.5in}
		\caption{\scriptsize (Left) Integrated $^{12}$CO $J=2-1$ emission toward G32.4$+$0.1 from $+39$ to $+47~\kms$.  Orange contours are NVSS $1.4~\text{GHz}$ continuum emission of the remnant.  Pixels with broadened molecular emission are shown as white squares.  We denote a comparison region of unbroadened, bright molecular emission with a larger red square.  (Right) The average spectrum from these pixels at the velocity of the BML identification (solid black) along with a spectrum of the unbroadened, bright comparison region averaged over the pixels denoted by the red square (dashed red).}\label{fig:G324+01}
\end{figure*}

G32.4$+$0.1 was discovered in the NRAO/VLA Sky Survey (NVSS) at $1.4~\text{GHz}$ \citep{condon+98}.  The remnant has clear shell-like structure visible in both radio and X-ray wavelengths \citep{yamaguchi+04}.  While the morphology of the remnant and its X-ray spectrum imply a strong synchrotron component, a radio spectral index has not yet been measured.  Of particular note are two lobes seen in nonthermal radio and X-ray emission along the eastern edge of the remnant \citep{ueno+05}.  These features appear to be local enhancements in nonthermal emission (\autoref{fig:G324+01}).  Moreover, they are adjacent to regions where the radio shell appears to break and little or no emission can be seen.  This morphology contrasts with the western edge where the radio shell appears more contiguous albeit no brighter than the eastern lobes.

We detect broadened molecular emission toward a cloud at a systemic velocity of $+43~\kms$ and roughly coincident with the eastern radio lobe of G32.4$+$0.1.  The associated MC extends along the southeastern and eastern part of the shell, although we detect velocity-broadening only along the northernmost portion of this cloud where the radio continuum of G32.4$+$0.1 is brightest.  The Gaussian velocity-width of molecular emission in this region ranges from $7.1$ to $8.3~\kms$.  

The association with the nonthermal enhancement toward G32.4$+$0.1 is highly suggestive, and the spatial correlation between the MC and the shell of the remnant imply an interaction scenario.  However, the kinematic distance inferred from this association is in tension with the values inferred from the $\Sigma-D$ relation ($18.5~\text{kpc}$) and X-ray \HI\ absorption ($22~\text{kpc}$) \citep{ueno+05}.  The $\Sigma-D$ estimate may be affected by the low flux density of G32.4$+$0.1 ($S_{1GHz} \sim 0.25~\text{Jy}$ from \citet{green14}), which renders surface brightness and size estimates uncertain.  Finally, it may be that the density of hydrogen toward this source is significantly larger than the $n_{H} = 1.0~\text{cm}^{-3}$ assumed for the latter estimate. 

\section{Discussion}\label{sec:dis}

\subsection{The SNR-MC Interaction Rate}\label{sec:rate}

\citet{rm01} argued that, because half of all SNRs are expected to result from Type II SNe and their progenitors are young, massive stars born in large MC complexes, it is expected that roughly half of SNRs should be in contact with MCs.  New information on the incidence of core-collapse SNe places additional constraints on the expected rate of this type of SNR-MC interaction.  Here, we discuss the total number of SNR-MC interactions in our galaxy as a fraction of all known Galactic SNRs, which we call the ``SNR-MC interaction rate.''  We examine this value in light of our findings and the factors that may contribute to an enhanced or reduced rate.

Recent transient surveys have greatly improved our understanding of the intrinsic fraction of both thermonuclear and core-collapse SN types in volume-limited samples.  For example, the Lick Observatory Supernova Search (LOSS) observed 175 SNe between $60$ and $80$ Mpc and determined that $57\%$ of SNe are Type II, another $19\%$ are Type Ib/c and the remaining $24\%$ are Type Ia \citep{llcf11}, implying that roughly $3/4$ of SNRs come from core-collapse SNe.  We might expect the SNR-MC interaction rate to be close to this value.

However, discounting the four objects in our sample mentioned in \autoref{sec:obs}, our fraction of candidate interactions (17/46 or $37\%$) is much lower than the observed fraction of core-collapse SNe.  If we rule out possible Type Ia SNe by looking only at SNRs with known pulsars or compact X-ray sources, which we infer resulted from core-collapse SNe, the total rate appears to change somewhat (7/15 or $47\%$).  We show in \autoref{sec:colines} below that virtually all SNR-MC interactions identified by other means also show broad CO and will be identified by our approach. However, in \autoref{sec:non} we illustrated the potential for false identifications. Therefore, 7/46 and 7/15 should be interpreted as upper limits.  The difference between these two rates also supports the hypothesis that SNR-MC interactions tend to be associated with core-collapse SNe, but both values are low compared to the expected rate.  There appears to be an underlying physical mechanism that suppresses the number of SNRs that interact with MCs compared to the known fraction of core-collapse SNe or the number of SNR-MC interactions detectable in CO lines.  

In \autoref{sec:colines}, we discuss the possibility of ongoing or recent interactions between SNRs and MCs that do not excite broadened $^{12}$CO $J=2-1$ emission.  We examine the possibility that some underlying physical mechanism suppresses the SNR-MC interaction rate in \autoref{sec:low} and in \autoref{sec:cons} we discuss the implications of a low SNR-MC interaction rate for sequential star formation and simulations of SN feedback.  

Finally in \autoref{sec:gamma}, we analyze the correlation between SNR-MC interactions and TeV gamma-ray emission from SNRs.

\subsection{To What Extent Are SNR-MC Interactions Detectable in $^{12}$CO $J=2-1$?}\label{sec:colines}

One of the most fundamental questions in our $^{12}$CO $J=2-1$ analysis of BML regions toward SNRs is what fraction of SNR-MC interactions are detectable in molecular emission?  \citet{jcws10} argue that OH maser detections are a reliable (if incomplete) signpost of SNR-MC interactions, so a test of the completeness of $^{12}$CO $J=2-1$ line broadening detections for SNR-MC interactions might be ``what fraction of SNRs with nearby OH maser emission have also been detected in broad-line molecular emission?''  In \autoref{tab:OH}, we list remnants with OH maser detections and whether molecular line studies, where they exist, support the presence of broadened molecular lines.

\begin{deluxetable}
{lllc}
\tabletypesize{\scriptsize}
\tablecaption{SNRs with Detected OH Maser Emission\label{tab:OH}}
\tablehead{
Cat. No. & Transitions in Previous Studies & LB? & Ref.\\
& &}
\startdata
G0.0$+$0.0	& CS (1$-$0, 5$-$4, 7$-$6)	&	Y	& 1,2\\
G1.05$-$0.1	&							&  		&	\\
G1.4$-$0.1	&							&  		&	\\
G5.4$-$1.2	& CO (1$-$0)				& N		& 3\\
G5.7$-$0.0	& CO (1$-$0)				& N		& 3\\
G6.4$-$0.1	& CO (1$-$0, 3$-$2)			& Y		& 4	\\
G8.7$-$0.1	&							&  		&	\\
G9.7$-$0.0	&							&  		&	\\
G16.7$+$0.1	& CO (1$-$0, 2$-$1)			& Y		& 5,6\\
G21.8$-$0.6	& CO (1$-$0), $^{13}$CO (1$-$0), C$^{18}$O (1$-$0), & Y 	& 7\\
&  HCO$^{+}$ (1$-$0) &&\\
G27.4$+$0.0	& CO (2$-$1)				& Y& 5 \\
G31.9$+$0.0	& CO (2$-$1), CS (2$-$1, 3$-$2, 5$-$4),	& Y		& 5,8 \\
& HCO$^{+}$ (1$-$0) &&\\
G32.8$-$0.1	& CO (1$-$0, 2$-$1) $^{13}$CO (1$-$0) & Y	& 9	\\
G34.7$-$0.4	& CO (2$-$1)				& Y		& 10 \\
G49.2$-$0.7	& CO (1$-$0, 2$-$1), $^{13}$CO (1$-$0),	& Y		& 11 \\
&  HCO$^{+}$ (1$-$0) &&\\
G189.1$+$3.0	& CO (1$-$0, 2$-$1)		& Y		& 5,12 \\
G337.0$-$0.1	&						&  		&	\\
G337.8$-$0.1	&						&  		&	\\
G346.6$-$0.2	&						& 		&	\\
G348.5$-$0.0	&						&  		&	\\
G348.5$+$0.1	& CO (1$-$0)			& Y		& 6	\\
G349.7$+$0.2	& CO (1$-$0)			& Y		& 13	\\
G357.7$-$0.1	& CO (1$-$0, 2$-$1, 4$-$3), & Y	& 14 \\
&  $^{13}$CO (1$-$0, 2$-$1), CS (2$-$1, 3$-$2), &&\\
&  HCO$^{+}$ (1$-$0), HCN (1$-$0), &&\\
& H$_{2}$CO (3$_{(2,2)}$$-$2$_{(2,1)}$, 3$_{(0,3)}$$-$2$_{(0,2)}$)& & \\
G357.7$+$0.3	&						&  &	\\
G359.1$-$0.5	& CO (1$-$0, 2$-$1), $^{13}$CO (1$-$0, 2$-$1), & ?	& 15 \\
& C$^{18}$O (1$-$0, 2$-$1), CS (2$-$1, 3$-$2),&&\\
&  HCO$^{+}$ (1$-$0), HCN (1$-$0), &&\\
& H$_{2}$CO (3$_{(2,2)}$$-$2$_{(2,1)}$, 3$_{(0,3)}$$-$2$_{(0,2)}$),&&\\
& SiO (2$-$1, 5$-$4) &&
\enddata
\tablecomments{SNRs are selected from those with identified OH maser emission features in \citet{frail96,green97,ygrr99,hyw08,hy09}.  For remnants where studies have also been performed in molecular line emission, we give the transitions observed.  We also indicate whether the study concluded that the observed emission exhibits line-broadening (LB) relative to the surrounding gas (Y) or there was no evidence of line broadening (N). For references that performed molecular line studies: (1) \citep{serabyn+92}; (2) \citep{tsuboi12}; (3) \citep{liszt09}; (4) \citep{atst99}; (5) this study; (6) \citep{rm00}; (7) \citep{zhou+09}; (8) \citep{rr99}; (9) \citep{zhou11}; (10) \citep{seta+98}; (11) \citep{km97}; (12) \citep{cornett77}; (13) \citep{rm01}; (14) \citep{lazendic+04}\tablenotemark{a}; (15) \citep{lazendic+02}\tablenotemark{b}}
\tablenotetext{a}{Only the observations mentioned in $^{12}$CO $J=4-3$ toward G357.7-0.1 have broad-line detections, although these data have not been published.}
\tablenotetext{b}{Broad-line absorption is seen in CO, CS, and HCO$^{+}$ while broad-line emission appears in $^{13}$CO.  See \autoref{sec:colines}.}
\end{deluxetable}

While all of the studies cited in \autoref{tab:OH} detect molecular features at roughly the same systemic velocity as the coincident OH maser emission, the incidence of broad-line detections is 13/15 or $87\%$.  Another study is inconclusive: \citet{lazendic+02} confirm the presence of shocked molecular hydrogen emission and broad-line absorption toward G359.1$-$0.5 but their results are ambiguous for the CO lines, so we have not included this object.

Molecular line broadening toward shocked MCs may not be detected for several reasons.  Molecular gas from a single cloud can be optically thick and obscure the shocked regions.  The faint line wings characteristic of broadened molecular lines may also be obscured by bright, narrow emission from other clouds.  These effects are usually more severe in lower energy transitions where line emission from quiescent clouds is intrinsically broad.  Indeed, both of the BML non-detections in \autoref{tab:OH} were performed in $^{12}$CO $J=1-0$ and toward SNRs near the Galactic Center (G5.4-1.2, G5.7-0.0) \citep{liszt09}.  Observations in higher energy transitions might show less emission from quiescent clouds along the line-of-sight while revealing broad-line emission.  Every observation we report that was performed in transitions at higher energies than $^{12}$CO $J=1-0$ has found signs of broadened molecular emission.

We infer that broad molecular line emission is a highly complete (e.g., $> 87\%$) tracer of SNR-MC interactions, and $^{12}$CO $J=2-1$ may exhibit broadened molecular lines in almost all instances of a SNR-shocked MC.  Detection of broadened $^{12}$CO $J=2-1$ lines may be a necessary, but not sufficient, condition for the presence of SNR-MC interactions.  This hypothesis would imply the limiting factor in assessing the ``true'' number of SNR-MC interactions in $^{12}$CO $J=2-1$ is not false negatives from optically thick or obscured emission, but false positives from multiple MCs in velocity-space that give the appearance of a broad line or from BML regions toward other sources such as \HII\ regions.  We report evidence of both effects, for example toward HB 3 where the coincident BML region appears to be associated with the nearby \HII\ region W3(OH), G54.1$+$0.3 where the line emission from multiple velocity components appears to have been mistaken for broad-line emission in previous work, and 4C$-$04.71 where BML regions are detected at multiple velocities implying contribution from other sources.

Given the high completeness we infer for detections of SNR-MC interactions observed in $^{12}$CO $J=2-1$, the ratio of BML regions we detected to observed SNRs (i.e., $37\%$) may only be an upper limit to the ``true'' SNR-MC interaction rate in the galaxy.  A lower SNR-MC interaction rate only exacerbates the discrepancy with the fraction of core-collapse SNe.  We infer that some additional mechanism is required in order to suppress the number of SNR-MC interactions occurring in the galaxy, and we explore possibilities and their consequences below. 
	
\subsection{Suppressing the SNR-MC Interaction Rate}\label{sec:low}

Suppression of the SNR-MC interaction rate relies on a SN event that is too distant from a MC for interaction to occur on a short timescale.  For core-collapse events, we propose that there are two main mechanisms by which suppression can occur.  One possibility is that massive stars are kicked or migrate away from their parent MC such that they are too distant to cause a SNR-MC interaction.  The second is a delay between the SN event and SNR-MC interaction, which we can express as some characteristic timescale or, equivalently, a characteristic separation between massive stars and MCs.  These mechanisms are not mutually exclusive, and indeed, the former might provide an explanation for the delay between SN explosions and SNR-MC interactions.  The evolution of the SNR-MC interaction rate with age provides a way of disentangling each possibility and its underlying cause.  

For example, one explanation of the low SNR-MC interaction rate may be a large, distinct population of runaway massive stars.  \citet{tnh11} detected 2547 runaway star candidates in Hipparcos, including over 200 stars with peculiar velocities from $10 - 120~\kms$ toward OB associations.  For a star migrating directly away from its parent cloud at $10~\kms$ over $10^{6}$ yr, a subsequent core-collapse SN would explode roughly $10$ pc from its starting position.  Thus, the time between explosion and SNR-MC interaction would be delayed by approximately $6000~\text{yr}$ (i.e., the age at which a SNR has a radius of $10~\text{pc}$).  

Age is a critical factor in the SNR-MC interaction rate.  We hypothesize that the SNR-MC interaction rate for SNRs of varying ages may be solely a function of the ``delay'' between the explosion and interaction.  This supposition is supported by the observation that the incidence of runaway stars is appreciable compared to the overall population of young, nearby stars.  \citet{tnh11} found that the runaway frequency of OB stars is $27\%$ for their sample.  This population of stars may explode as SNe and interact with MCs only as evolved SNRs or never exhibit signatures of SNR-MC interactions.

Assuming massive stars are ``kicked'' or otherwise migrate a large distance from their parent MC, an analysis of the SNR-MC interaction rate with SNR age would provide a constraint on the delay caused by this migration.  A constant SNR-MC interaction rate with age would suggest that there are two distinct populations of massive stars; those that remain close to their parent MC and those that migrate a large distance such that all known Galactic SNRs are too young for the latter to have encountered a MC.  The more likely scenario is that the SNR-MC interaction rate increases with SNR age, which suggests that massive stars migrate a short distance (i.e., $< 30~\text{pc}$) from their parent MC over their lifetimes.  We would expect the SNR-MC interaction rate to level off at a certain SNR age.  This analysis may provide a constraint on the distribution of massive star runaway velocities and the fraction of massive stars that are runaways.

We note several SNRs with age determinations from free-expansion velocity or pulsar timing (\autoref{sec:anc}).  However, the number of such SNRs is too low to draw any statistically meaningful conclusions.  As more SNR-MC interactions are identified, accurate age determinatons may be obtained from kinematic distances and modeling of SNR size, morphology, and surface brightness.  \citet{vink12} provides an analytic overview of age determinations that accounts for SN parameters as well as momentum and radiative loss.  Once ages have been determined for a sufficient number of SNRs, an estimate of the dependence of the SNR-MC interaction rate with age should be possible.

\subsection{Astrophysical Implications of a Low SNR-MC Interaction Rate}\label{sec:cons}

Another rich field of study is the direct impact SNRs may have on star formation.  Models of sequential or triggered star formation rely on an understanding of the feedback between massive stars and star forming regions.  While the relationship between OB associations and star-forming regions has long been recognized \citep{el77,ha77}, the contribution of shocks from SNRs is more ambiguous.  Some studies have proposed that these shocks will compress molecular regions and accelerate triggered star formation \citep{thi81}, although it is possible the shock velocities involved are dissociative and will destroy the molecular gas instead.  There is also evidence that SNRs play little or no role in regulating star formation in individual star clusters due to the time delay before SNe occur, but turbulent outflows powered by SNe may suppress star formation in giant MCs or even galaxy-wide star formation \citep{kbad14}.  

Recent numerical simulations of SN feedback have examined the impact SN feedback has on galactic environments in order to evaluate the importance of SNRs in star formation \citep[e.g.,][]{hopkins12,agertz13,aumer13,stinson13,hopkins14}.  Many of these simulations fail to produce the observed stellar masses due to excessive cooling or the lack of momentum transfer from SNe to giant MCs.  This problem may be exacerbated by a large fraction of SNe that explode far from any molecular gas.  Over short timescales, a large fraction of SNe may play no role in galaxy-wide feedback.

In the absence of momentum transfer from SNRs, it is possible that stellar feedback mechanisms other than SNe are dominant.  For example, \HII\ regions are known to photoionize dense MCs and may suppress star formation where a sufficient level of background radiation is present.  This hypothesis is supported by observations of giant \HII\ regions (such as 30 Doradus in the LMC) indicating that radiation pressure from stars is the dominant mechanism in adding energy and momentum to the surrounding region \citep{lopez+11}.  The relative role of each mechanism must be assessed in the context of specific timescales and physical conditions.  Our observations support arguments that SNRs may contribute less to stellar feedback than previously thought.

\subsection{Gamma-Ray Emission from SNR-MC Interactions}\label{sec:gamma}

Particle acceleration is one of the most important galactic processes in which SNRs play a central role.  Cosmic rays are produced in SNRs where the SN explosion can deliver enough energy to rapidly accelerate nuclear particles \citep{berezhko+96}, and the energy density of cosmic rays produced in the shock region increases with the density of the surrounding medium \citep{jj99}.  Enhanced turbulence and magnetic fields can excite significant hadronic particle acceleration which produces $\pi^{0}$ mesons that decay immediately into gamma-rays.  Consistent with this hypothesis, there are a number of instances where TeV gamma-rays are reported from regions of interaction between SNRs and MCs \citep[e.g.,][]{aharonian+06b,albert+07a,acciari+09}.

A challenge to such studies is the complex and dense population of MCs and the high density of TeV gamma-ray sources in the inner Milky Way where most known TeV sources lie. There may also be a tendency to target regions of possible interaction in searching for gamma-ray sources, potentially introducing a bias. These issues can be mitigated in our survey. We use the positions of the specific SNRs targeted in the CO survey as priors to identify possible TeV counterparts, which we accept if their positions coincide with the SNRs within $10'$. This positional tolerance was adopted from a conservative estimate of the positional accuracy of $6'$ for the TeV measurements \citep[e.g.,][Felix Aharonian, private communication]{hess+04}, plus an allowance for the extent of the SNR. We can then compare the number of detections for the SNRs where we detect broad CO lines indicative of SNR-MC interactions with the number where we do not find broad lines. We adopt the conventional threshold of $5\sigma $ for a TeV detection.  \citet{hahn14} has evaluated the HESS Milky Way survey (the source of nearly all the TeV detections we use) and shown that all of the sources at this detection level have been identified, that is, the sample is complete to this level of significance. We correct the detection rates for chance coincidences by identifying TeV sources with the same positional tolerance relative to 120 positions selected arbitrarily in the same region. 

Where there were possible identifications, we report the separation between the approximate location of the SNR-MC interaction region or the center of the SNR for remnants without SNR-MC interactions and the nearest TeV gamma-ray source in \autoref{tab:gamma}. We have taken positions for the TeV sources from TeVCat \citep{wh08}. There are nine cases within our positional criteria, four with indications of MC interactions and five without. Our test on arbitrary positions indicates that there are about two false identifications, one for the cases with interactions and one for those without. That is, there are three true cases coinciding with interactions and four not coinciding. We have estimated the number of SNRs with TeV measurements from \citet{bochow11}, as shown in \autoref{tab:survey}. The two sample sizes are virtually the same (19 without interactions, 20 with), so the rates of detections are identical within statistical errors.

\begin{deluxetable*}
{lcccccl}
\tabletypesize{\scriptsize}
\tablecaption{SNRs with Possible TeV Gamma-Ray Emission at $5\sigma$ Signifiance\label{tab:gamma}}
\tablehead{
Cat. No. & $\alpha$ (J2000) & $\delta$ (J2000) & BML Detection & $\gamma$-ray Separation\tablenotemark{a}
 & $\gamma$-ray Source & \parbox[c]{1.34in}{Ref.}\\
& (h m s) & (d m s) & & arcmin & & }
\startdata
G08.3$-$0.0  &		$18~04~28$ &	$-21~47~00$ &	D	&	5.1 & HESS J1804-216 & \citet{aharonian+06b} 	\\ %
G11.2$-$0.3  &		$18~11~25$ &	$-19~23~20$ &	D	&	22  & HESS J1809-193 & \citet{renaud+08} 		\\ 
G12.8$-$0.0  & 		$18~13~37$ & 	$-17~49~00$	&		&	1.4	& HESS J1813-178 & \citet{aharonian+06b} 	\\ 
G17.0$-$0.0  & 		$18~21~57$ & 	$-14~08~00$	&		&	57	& HESS J1825-137 & \citet{aharonian+06c}	\\ %
G18.1$-$0.1  & 		$18~24~34$ & 	$-13~11~00$	&		&	42	& HESS J1825-137 & \citet{aharonian+06c}	\\ %
G21.5$-$0.9  & 		$18~30~50$ & 	$-10~09~00$	&		&	17	& HESS J1831-098 & \citet{sheidaei11}		\\ 
G27.4$+$0.0  &		$18~41~18$ &	$-04~54~20$ &	D	&	39  & HESS J1841-055 & \citet{aharonian+08b} 	\\ 
G29.7$-$0.3  &		$18~46~26$ &	$-02~57~30$ &	D	&	1.5 & HESS J1846-029 & \citet{djannati-atai+08} \\ 
G74.9$+$1.2	 &		$20~16~02$ & 	$+37~12~00$	&		&	0.1	& VER J2016+371	 & \citet{aliu11}			\\
G111.7$-$2.1 &		$23~24~15$ &	$+58~48~20$ &	D	&	7.9 & TeV J2323+588  & \citet{becker+91} 		\\ 
G120.1$+$1.4 & 		$00~25~18$ & 	$+64~09~00$	&		&	2.1	& TeV J0025+641	 & \citet{acciari+11} 		\\
G130.7$+$3.1 &		$02~05~41$ & 	$+64~49~00$ &		&	2.0	& TeV J0209+648	 & \citet{aleksic+14}		\\
G184.6$-$5.8 & 		$05~34~31$ & 	$+22~01~00$	&		&	0.2	& TeV J0534+220  & \citet{aharonian+06} 	\\ 
G189.1$+$3.0 &		$06~17~24$ &	$+22~26~20$ &	D	&	8.4 & TeV J0616+255  & \citet{acciari+09} 		 
\enddata
\tablecomments{SNRs with nearby TeV gamma-ray detections at the $5\sigma$ significance level.  We report the position of the detected BML region from our study for SNRs with BML detections or the location of the SNR itself for those without BML detections from \citet{green14}.  The gamma-ray separation is determined from the difference between the reported position and the nearby TeV gamma-ray source.  The gamma-ray source identifiers are also given with the reference for the position of each source indicated.}
\tablenotetext{a}{Only cases with separation $< 10'$ are accepted for our study.}
\end{deluxetable*}

The similar number of associations between SNRs and gamma-ray emission in these two samples does not support a strong correlation between SNR-MC interactions and TeV gamma-ray emission. The observed gamma-ray detections may also be contaminated by particle acceleration in pulsars, which are known gamma-ray sources. For example, the Crab pulsar has long been associated with hadronic particle acceleration and TeV gamma-ray emission \citep{hayakawa58,fazio+68,og69}.  TeV gamma-ray surveys typically distinguish between ``compact'' and ``extended''' gamma-ray sources, which may aid in resolving the ambiguity between gamma-rays from pulsars and more extended regions along the SNR shell.  However, this distinction may fail to separate pulsar point sources from SNR-MC interactions where the conditions for particle acceleration are met only in a small region.  Moreover, \HI\ clouds are observed near SNRs with a high enough density gradient to excite particle acceleration at shock fronts \citep{fukui+12,xing+15}. In the absence of a pulsar, it may not be necessary to have a SNR-MC interaction to create a gamma-ray source toward a SNR, although the contact discontinuity between SNR ejecta and the densest parts of the ISM would still be expected to produce the brightest and hardest gamma-ray sources.

We can account for these uncertainties to an extent.  Ten out of the fourteen SNRs we report with TeV gamma-ray emission at $5\sigma$ significance have pulsars or coincident, compact X-ray sources, including five of the six objects with BML regions and five of the eight without. Removing these sources leaves only G08.3$-$0.0 and Tycho as the two objects with TeV gamma-ray sources detected at close separations, and while we report a coincident BML region toward the former, we do not detect any strong $^{12}$CO emission toward the latter. G08.3$-$0.0 is a recently discovered object (\autoref{sec:G083-00}) and it is possible that an undiscovered pulsar exists in the SNR and provides the particle acceleration necessary to generate TeV gamma-rays. \citet{higashi+08} discuss several possible sources for the gamma-ray detection in this region, including the pulsar PSR B1800-21, although this source appears to be separated from the gamma-ray emission by at least $10'$. Barring these possibilities, G08.3$-$0.0 represents the best candidate in our sample for TeV gamma-ray emission from a SNR-MC interaction.

In addition to G08.3$-$0.0, compelling evidence for TeV gamma-ray emission can be seen toward a number of SNR-MC interactions, notably W28 \citep{aharonian+08c} and IC 443 \citep{acciari+09}.  Although these cases show that SNR-MC interactions can produce TeV sources in some cases, the lack of a strong correlation in our study shows that other processes must play an important, perhaps dominant, role. Larger statistical samples are needed to evaluate the association of TeV sources with SNR-MC interactions more definitively.

\citet{bochow11} lists all measurements of SNRs, detections and upper limits from HESS. Out of the nineteen SNRs we observed with evidence for SNR-MC interaction, twelve also have evidence for gamma-ray emission at the $2\sigma $ significance level including six of the nine SNRs with newly reported interactions. This fraction is appreciably higher than the overall incidence of gamma-ray emission toward SNRs, which is 24/50 in our sample. Previous systematic studies for gamma-ray emission toward SNRs have found an even lower ratio of SNRs with coincident TeV gamma-ray emission \citep[1/3 in][]{bochow11}. At this level, our findings seem to support the hypothesis that TeV gamma-ray emission is correlated with SNR-MC interactions. However, confirmation of gamma-ray emission at this level is problematic. Detections at $2\sigma $ may be close to the confusion limit with background sources along the Galactic plane. In addition, the diffuse emission of the Galaxy can contribute to such detections without being associated with a specific source. Combined with the positional uncertainty in gamma-ray experiments, the correlation between SNR-MC interactions and TeV sources at $2\sigma $ is tantalizing but not necessarily reliable. 

We conclude that there is no convincing evidence with the existing data that SNR-MC interactions are a dominant source of TeV gamma-ray emission from SNRs. Other sources of particle acceleration in SNRs, especially pulsars and interaction with the \HI\ gas in the cold neutral medium, may contribute significantly to the number of gamma-ray detections.  

\section{Conclusion}

Our survey of $^{12}$CO $J=2-1$ molecular spectroscopy toward 50 SNRs is the first large, systematic search for this shock tracer toward SNRs.  We demonstrate that our technique for analyzing $^{12}$CO $J=2-1$ data cubes satisfactorily reproduces broad-line detections in systems with known BML features.  In applying this technique to our data set, we have found:

\begin{enumerate}

\item BML features were detected toward nine remnants (G16.7$+$0.1, Kes 75, 3C 391, Kes 79, 3C 396, 3C 397, W49B, Cas A, HB 3, IC 443) with previous evidence of SNR-MC interactions and nine remnants (G08.3$-$0.0, G09.9$-$0.8, G11.2$-$0.3, G12.2$+$0.3, G18.6$-$0.2, G23.6$+$0.3, 4C$-$04.71, G29.6$+$0.1, G32.4$+$0.1) with no previous evidence of interaction.  The BML region detected toward HB 3 is likely associated with the \HII\ region W3(OH).  We are uncertain whether the BML regions associated with 3C 396 and W49B represent SNR-MC interactions, and detailed follow-up toward these clouds is required to draw any firm conclusions.  For one of the remaining remnants, G54.1$+$0.3, we find that previous indications of a SNR-MC interaction may have been due to the blending of multiple narrow line components.

\item From our analysis of SNRs with signs of molecular shocks from OH masers, we infer broadened $^{12}$CO $J=2-1$ features are a necessary but not sufficient condition to determine the presence of SNR-MC interactions.  Confirmation must come from other sources, such as $^{12}$CO line ratios, to demonstrate that the gas in BML regions is physically distinct from the surrounding material.  The total number of SNR-MC interactions we observed as a fraction of observed SNRs may only be an upper limit to the ``true'' SNR-MC interaction rate in the galaxy.

\item The fraction of detected BML regions toward SNRs with known pulsars or compact X-ray sources (7/15) is slightly larger than the overall number (17/46), supporting the hypothesis that SNR-MC interactions are associated with core-collapse SNe from massive stars.  However, the number of BML regions we detect toward SNRs is significantly lower than the incidence of core-collapse SNe.  We infer that age may be a critical factor in the overall rate of SNR-MC interactions.  A low SNR-MC interaction rate may also have far-reaching consequences for models of stellar feedback and sequential star formation.

\item We found no convincing evidence for a correlation between TeV gamma-ray sources and SNR-MC interactions. The number of TeV detections within $10'$ of SNRs with BMLs is the same as the number for SNRs with no evidence for interactions. We conclude that SNR-MC interactions are not a dominant source of TeV gamma-rays in the galaxy. As an alternative, pulsars or interaction with \HI\ gas in the cold neutral medium may contribute significantly to the detection rates for TeV gamma-rays toward SNRs.

\end{enumerate}

\acknowledgments

The Heinrich Hertz Submillimeter Telescope is operated by the Arizona Radio Observatory with partial support from National Science Foundation grant AST 1140030.  Funding for this research was provided by NASA through Contract Number 1255094 issued by JPL/Caltech.  We would like to express our gratitude to Crystal Brogan for providing her data from BGGKL06 for remnants used in this study as well as David Moffett for providing 20 cm radio continuum data of 3C 391.  We also thank Joachim Hahn for his helpful comments.

\appendix

\section{Individual SNR Distance Determinations}\label{sec:altdist}

Kepler (G4.5$+$6.8) is a young SNR whose proper motion is fast enough to be measured.  From \textit{Chandra} ACIS-S measurements of Kepler's expansion and a shock velocity inferred from X-ray synchrotron emission, \citet{vink08} determined that Kepler is located at a distance of $4.0$ kpc.

G11.2$-$0.3 was observed in neutral hydrogen by \citet{becker+85} who found that weak \HI\ absorption can be seen up to $55~\kms$ and roughly consistent with a distance of $4.4$ kpc.

G12.8$-$0.0 has an association with stellar cluster Cl $1813−178$ whose distance is known to be $4.8$ kpc \citep{mdfk11}.

G15.9$+$0.0 does not yet have a trustworthy distance estimate, although multiple $\Sigma$-D relationships have been calculated for this remnant.  \citet{caswell+82} calculated the distance to be $16.7$ kpc \citep[updated to $8.5$ kpc for the correct galactocentric distance as in][]{reynolds+06}.  \citet{pdvu14} use the $\Sigma$-D relationship and place the remnant slightly further at $10.4$ kpc.  We use $8.5$ kpc in our analysis for consistency with previous studies.

G16.7$+$0.1 has an associated pulsar with high Galactic \HI\ column density along its line-of-sight \citep{hag03}, roughly consistent with a distance of $10$ kpc.

It has been argued that the complex of \HII\ regions near G18.1$-$0.1 implies an association with the SNR.  Several studies have used \HI\ absorption estimates and inferred a distance of $\sim 4.0$ kpc to G18.1$-$0.1 \citep[e.g.,][]{downes+80,kolpak+03,anderson+09,paron+13}. However, new \HI\ absorption estimates to the remnant itself imply this assumption may be in error and the remnant is located beyond the \HII\ regions at $5.6$ kpc \citep{leahy+14}, which we use in our analysis. 

G21.5$-$0.9 has been identified from \HI\ absorption between systemic velocities $+67~\kms$ and $+69~\kms$ for a distance of $4.8$ kpc \citep{tl08}.

Radio recombination lines at systemic velocities from $+42$ to $+99~\kms$ provide kinematic constraints on the distance to G29.6$+$0.1 \citep{lockman+96}.  These values, along with the distance to the nearby pulsar AX J1845-0258 \citep{torii+98}, led \citet{gaensler+99} to estimate that G29.6$+$0.1 is at a distance of $10$ kpc.

\citet{tian+08a} measured the distance to G27.4$+$0.0 (4C-04.71) in \HI\ absorption to be $8.5$ kpc.

Based on X-ray absorption measurements, \citet{vgtg00} determined that the distance to G29.6$+$0.1 is $5--15$ kpc (i.e., $10 \pm 5$ kpc).

G29.7$-$0.3 (Kes 75) has several distance estimates derived from \HI\ absorption, separate $\Sigma-D$ relationships, and CO associations \citep[e.g.,][]{caswell+75,milne79,bh84,scyk09}.  We adopt the distance $6.0$ kpc, derived by \citet{leahy+08a} from \HI\ absorption studies.

\citet{case+98} derived a distance of $12.9$ kpc to G31.5$-$0.6 based on $\Sigma-D$ estimates, and this value has been used in previous studies of this remnant \citep[e.g.,][]{mavromatakis+01}.  We adopt this distance in our study.

The distance to G32.4$+$0.1 was measured in X-ray absorption by \citet{yamaguchi+04}, who determined the remnant lies at $d = 17$ kpc.

\citep{frail+89,gd92} both measured the distance to G33.6$+$0.1 (Kes 79) kinematically from \HI\ and OH absorption, respectively.  Their measurements both agree with a distance $7.1$ kpc.

The distance to 3C 396 has been derived in several ways, sometimes with discrepant results.  \citet{caswell+75} use \HI\ absorption to determine 3C 396 is $> 7.7$ kpc, while X-ray \HI\ column density and CO associations indicate $6.2 - 8$ kpc \citep{okad03,hrar09,scyk11}.  We adopt the latest kinematic estimate of $6.2$ kpc as in \citet{scyk11}.

Toward 3C 397, \HI\ absorption indicates a distance of at least $7.5$ kpc while kinematic estimates from CO give $10.3$ kpc \citep{caswell+75,jcws10}.  We use this estimate in our analysis.

Enhanced \HI\ absorption toward W49B suggests the distance is $12.5--14$ kpc \citep{lockhart+78,brogan+01}, from which we adopt the lower estimate of $12.5$ kpc.

\citep{leahy+08b} determine the distance to G54.1$+$0.3 from morphological association with CO around $53~\kms$ to be $6.2$ kpc.

Optical observations of G59.8$+$1.2 reveal line emission in H$\alpha$, H$\beta$, [N II], [S II], and [O III] \citep{bmxa05}.  These data suggest that the SNR has a large angular size ($\sim 20'$) and slow shocks ($< 70~\kms$).  G59.8$+$1.2 is also obscured by a \HI\ column density of $N_{\text{H}} \sim 2.5 - 2.8 \times 10^{22}$ cm$^{-2}$.  For a Galactic longitude of $l = 59.8^{\circ}$ and a galactocentric distance of the Sun of $R_{0} = 8$ kpc, the minimum galactocentric distance along a line-of-sight near the mid-plane is approximately $7$ kpc.  We can apply an estimate of the distribution of neutral hydrogen along the line-of-sight to G59.8$+$1.2 determined in \citet{kd08} to infer a distance to this SNR.  For $n_{0} = 0.9$ cm$^{-3}$, $R_{n} = 3.15$ kpc and assuming the galactocentric distance is $7$ kpc $\lesssim r \lesssim 35$ kpc, the average mid-plane volume density of neutral hydrogen is

\begin{equation}
n_{H}(r) = n_{0} e^{-(r - R_{0})/R_{n}}
\end{equation}

\noindent The distance to which the neutral hydrogen density profile integrates to the measured value of $2.6 \times 10^{22}~\text{cm}^{-2}$ toward G59.8$+$1.2 is $d = 7.3$ kpc, which we use as the distance to this SNR.

Associated \HI\ and CO features toward G63.7$+$1.1 yield a distance of $3.8$ kpc \citep{wlt97}.

The distance to G69.0$+$2.7 (CTB 80) has been determined from \HI\ emission to be approximately $1.5$ kpc \citep{lr12, pkgk13}.

\HI\ absorption and a possible CO association toward G74.9$+$1.2 (CTB 87) yield a distance estimate of $6.1$ kpc \citep{krfb03}.

G76.9$+$1.0 has an associated pulsar whose coordinates and spectra best place the SNR in the Outer Arm at $d \sim 10$ kpc \citep{agrs11}.

\citet{leahy+12} determined the distance to G84.2$-$0.8 from X-ray absorption to be $6$ kpc.

Proper motion measurements of Cas A along with spectroscopic measurements of the expansion velocity suggest the remnant is located at a distance of $3.4$ kpc \citep{rhfw95}.

\citet{hailey+94} determined the distance to CTB 1 kinematically from optical spectroscopy to be $3.1$ kpc.

The distance to Tycho has historically been controversial \citep[e.g.,][]{chevalier+80,schwarz+80}.  Measurements of \HI\ absorption at the near and far side of Tycho suggest the remnant is located at $d=$ 2.5--3.0 kpc \citep{tian+11} while X-ray proper motion studies suggest $4\pm1$ kpc \citep{hayato+10}.  We adopt the near distance estimate $2.5$ kpc in agreement with models of the shock wave \citep{smith+91,ghavamian+01} and measurements of a reported companion star \citep{ruiz-lapuente+04}.

Several observations of G130.7$+$3.1 (3C 58) in \HI\ absorption indicate a distance of $2 - 3$ kpc \citep{gg82,rgkh93,wlt94,kothes13}, from which we adopt the latest estimate of $2$ kpc.

As we indicate in \autoref{sec:known}, HB 3 is located near W3(OH) in the W3 star-forming region.  \citet{hachisuka+06} measure the distance to W3(OH) kinematically from water masers and find it is located at $d = 2$ kpc.

Proper motions of ejecta toward G184.6$-$5.8 (Crab/Crab Nebula) in optical and radio wavelengths yield a distance of $2$ kpc \citep{fs93,bhfb04,rfy08}.

\citet{welsh+02} constrained the distance to IC 443 by examining stars whose spectra indicated Na\I\ and Ca\II\ absorption from the remnant.  These stars constrained bracketed the distance at $1.5$ kpc.

The remaining three SNRs -- G12.5$+$0.2, G23.6$+$0.3, and G30.7$-$2.0 -- have no distance estimates and either have no known type or are classified as having composite morphology (\autoref{tab:survey}).  However, the targets have observed sizes and $1~\text{GHz}$ surface brightnesses, to which we can apply the $\Sigma - D$ relation and obtain a distance approximation.  We acknowledge that this relationship should not be applied to remnants that are not shell-like unless the shell structure is well-resolved or bright enough that all other emission can be ignored.  Therefore, these distance values should not be interpreted as a measurement to each SNR but a starting point we calculate for the purposes of our analysis.  Using the relationship calculated in \citet{pdvu14} and the angular size and $1~GHz$ flux in \citet{green14}, we find distances of $15$ kpc to G12.5$+$0.2, $6.9$ kpc to G23.6$+$0.3, and $8.7$ kpc to G30.7$-$2.0.

\section{Ancillary Data on Previously Detected BML Regions Toward SNRs}\label{sec:anc}

\textit{G16.7$+$0.1} is a composite remnant initially identified at $5~\text{GHz}$ in \citet{helfand89}.  The remnant has a steep spectral index between $1.4~\text{GHz}$ and $5~\text{GHz}$ of $\alpha \sim -0.6$.   Multiple OH $1720~\text{MHz}$ maser searches have revealed a source along the southern half of the remnant, consistent with one of the brightest regions of the SNR at $5~\text{GHz}$ \citep{green97,hyw08}.  This maser emission occurs at $+20~\kms$, and a tentative detection of OH (1667 MHz) absorption has also been reported at $+25~\kms$.

\textit{Kes 75} was initially identified as a SNR (referred to as G$29.7-0.2$) between 408 and $5000~\text{MHz}$ by its spectral index \citep{sg70}.  The spectral index measured in that study demonstrated nonthermal emission with $\alpha=-0.60$.  Low-frequency observations of this SNR are dominated by the outer shell where the emission demonstrates has an even steeper spectral index, up to $\alpha=-0.8$ \citep{kassim92}.  While the overall morphology appears to be composite, the steep spectral indices, even at high frequencies, suggests that the shell-like structure may be the dominant source of nonthermal emission. An X-ray pulsar was identified toward Kes 75 by \citet{gvbt00}.  Timing measurements suggest that the pulsar is relatively young at $723~\text{yr}$.

\textit{Kes 79} was first confirmed as a SNR at 408 and $5000~\text{MHz}$ by \citet{caswell+75}.  Subsequent pulsar detection and X-ray timing observations were performed with XMM-Newton \citep{ghs05,hgcs07}.  These observations confirmed the existence of a 105 ms X-ray pulsar toward Kes 79; however, the characteristic age was determined to be $\tau_{c} = 2P/\dot{P} > 8~\text{Myr}$.  This value is significantly larger than the age estimated in \citet{sun04}, where the authors identified a bright inner ring of material at $0.5 - 3~\text{keV}$ as a wind-driven shell and, assuming this material is expanding outwards at $5000 - 10000~\kms$, inferred an age of $\sim 6000$ yr.  Thus, while Kes 79 is likely a shell-like SNR with an embedded pulsar, the timing measurements imply the pulsar was spun up at birth and the younger age is more accurate.

OH lines at 1666, 1667, and $1720~\text{MHz}$ have been seen in both emission and absorption toward Kes 79 \citep{green89,green97,kfgc98,swdf03}.  The OH $1720~\text{MHz}$ maser was observed at $+90~\kms$ in a single-dish detection, although the source has yet to be localized.  In absorption, OH has been observed at $+12$ and $+55~\kms$ and in a broad feature from $+95$ to $+115~\kms$.  It is inferred that this last feature corresponds to a coincident MC, and the broadening in OH is due to a shock interaction with the SNR.

\textit{3C 396} was classified as a SNR by \citet{sg70}. 20 cm emission maps reveal composite morphology with significant brightening along the western half of the remnant, and especially toward the southwest \citep{patnaik+90}.  The spectrum at this frequency appears nonthermal ($\alpha = -0.42$), and measurements of the spectral index at 90 cm agree with this value \citep{kassim92}.  \citet{green97} reported detection of OH $1720~\text{MHz}$ maser emission at more than one pointing in single-dish observations of the remnant, although no velocity information or localization was obtained from these observations.  This detection was subsequently confirmed by \citet{kfgc98} in VLA observations, and the authors report the presence of a $+70~\kms$ OH line.

\textit{3C 397} was resolved and distinguished from a nearby \HII\ region to its west by \citet{kv74}.  The remnant is bright and compact, with several enhancements in the nonthermal radio emission to its west.  The spectral index is steep from 6 to 20 cm in this region ($\alpha = -0.66$) relative to the rest of the shell ($\alpha = -0.59$) \citep{ar93}.  Observations at $1720~\text{MHz}$ revealed no OH maser emission and imply that if there is a SNR-MC interaction the physical conditions may be too extreme to excite a maser \citep{green97}.

\textit{W49B} was initially identified as a strong radio source and probable SNR at $408~\text{MHz}$ by \citet{sg70}.  The remnant radio spectrum has a moderately nonthermal spectral index between $408~\text{MHz}$ and 1$5~\text{GHz}$ of $\alpha \sim -0.48$ \citep{gbl75}.  Thus far, no OH maser emission has been detected toward this remnant in targeted searches \citep[e.g.,][]{green97}.

\textit{G54.1$+$0.3} is a diffuse, low surface brightness radio source identified as a SNR in \citet{reich85}. The source is significantly polarized ($7\%$) at 4.7$5~\text{GHz}$, but has a flat spectral index at this frequency of $\alpha \sim -0.1$. Subsequent observations in the X-ray with ASCA and at $1175~\text{MHz}$ with Arecibo revealed a 136 ms pulsar, which combined with the SNR morphology, suggests that G54.1$+$0.3 is a Crab-like remnant with an age of $\sim 2900~\text{yr}$ \citep{clbg02}.

\textit{HB 3} is a large SNR with well-defined shell structure at $2.7~\text{GHz}$ \citep{vk74}.  The remnant is embedded in a star-forming region that contains the nearby \HII\ region W3.  Spectral indices from $38~\text{MHz}$ to $2.7~\text{GHz}$ ($\alpha \sim -0.5$) suggest HB 3 has a strong nonthermal component.  There is also evidence for a SNR-MC interaction from several OH $1720~\text{MHz}$ masers that have been detected to the southeast of this SNR and near W3 \citep{kfgc98}.  One of these detections, referred to as ``W3 northwest'' in \citet{kfgc98}, occurs at $-38.6~\kms$ and is directly to the southeast of the remnant.  Toward HB 3, this systemic velocity corresponds to a distance of $3.7~\text{kpc}$, slightly farther than inferred from this $\Sigma-D$ relation, although HB 3 is extended ($\sim 80'$) and presumably an old remnant at this distance ($\sim 30,000~\text{yr}$ in \citet{lazendic+06}).  Thus, surface brightness and size estimates are less reliable than for younger remnants and may yield discrepant distances compared to other methods.   


\clearpage

\end{document}